%% file: aggeffort.tex
\documentclass[12pt]{article}

\usepackage[english]{babel}
\usepackage[utf8x]{inputenc}
\usepackage[T1]{fontenc}

\usepackage[a4paper,margin = 1in]{geometry}
\usepackage{ifthen}
\usepackage{amssymb}
\usepackage{tikz}
\usepackage{amsthm}
\usepackage{amsmath}
\usepackage{mathtools}
\usepackage{graphicx}
\usepackage[colorinlistoftodos]{todonotes}
\usepackage[colorlinks=true, allcolors=blue]{hyperref}
\usepackage{float}
\usepackage{enumitem}
\usepackage{xcolor}
\usepackage{comment}
\usepackage{algorithm}
\usepackage{algorithmic}
\usepackage[nameinlink]{cleveref}
\crefname{equation}{}{}
\crefname{example}{Example}{}
\crefname{table}{Table}{}
\crefname{section}{Section}{}
\crefname{theorem}{Theorem}{theorems}
\crefname{lemma}{Lemma}{theorems}
\crefname{corollary}{Corollary}{}
\crefname{figure}{Figure}{}
\crefname{algorithm}{Algorithm}{}
\usetikzlibrary{automata,positioning}
\usepackage[bottom]{footmisc}

\newcommand{\RNum}[1]{\uppercase\expandafter{\romannumeral #1\relax}}

\newcommand{\C}{\mathcal{C}}
\newcommand{\B}{\mathcal{B}}

\newcommand*{\rom}[1]{\expandafter\@slowromancap\romannumeral #1@}

\input{macros}

\input{macros-math}

\def\estar{e^*}

\title{\bf Aggregate Play and Welfare in Strategic Interaction on Networks}
\author{
  Karan N. Chadha \thanks{Karan is with the Electrical Engineering Department at the Indian Institute of Technology Bombay, Mumbai, India 400076. Email : karanchadhaiitb@gmail.com}\\
  \and
  Ankur A. Kulkarni \thanks{Ankur is with the Systems and Control Engineering group at the Indian Institute of Technology Bombay, Mumbai, India 400076. Email : kulkarni.ankur@iitb.ac.in}\\
}
\date{\today}

\begin{document}

\maketitle










\begin{abstract}
In recent work by Bramoull\'{e} and Kranton~\cite{bramoulle2007public}, a model for the provision of public goods on a network was presented and relations between equilibria of such a game and properties of the network were established. This model was further extended to include games with imperfect substitutability in Bramoull\'{e} \etal~\cite{bramoulle2014strategic}. The  vast multiplicity of equilibria in such games along with the drastic changes in equilibria with small changes in network structure, makes it challenging for a system planner to estimate the maximum social welfare of such a game or to devise interventions that enhance this welfare. Our main results address this challenge by providing close approximations to the maximum social welfare and the maximum aggregate play in terms of only network characteristics such as the maximum degree and independence number. For the special case when the underlying network is a tree, we derive formulae which use only the number of nodes and their degrees. These results allow a system planner to assess aggregate outcomes and design interventions for the game, directly from the underlying graph structure, without enumerating all equilibria of the game, thereby significantly simplifying the planner's problem. A part of our results can be viewed as a logical extension of \cite{pandit2018refinement} where the maximum weighted aggregate effort of the model in \cite{bramoulle2007public} was characterized as the weighted independence number of the graph.
\end{abstract}




\section{Introduction}
In many settings of competition, the actions of only a subset of all agents directly affect the utility of an  agent. The underlying structure of these interactions can be  captured using a network or graph. The study of the resulting \textit{strategic interaction on a network} provides insights into how interconnections, such as physical connections or economic and social interdependencies, affect the  outcomes of interactions. These models have been applied for understanding diverse contexts such as crime, segregation, investment, research cooperation and so on. 

In this paper, we consider noncooperative games of positive externalities where players are situated on nodes of a network whose edges represent preexisting interconnections through which the actions of the neighbors of an agent influence the latter's utility. 
Beginning with the work of Bramoull\'{e} and Kranton~\cite{bramoulle2007public} on the provision of local public goods in the presence of networks, and extending further to Bramoull\'{e} \etal~\cite{bramoulle2014strategic}, many beautiful results have been found relating the nature of equilibria in such games to network structure. For instance, in the particular case of a public goods game on a network, assuming perfect substitutability, Bramoull\'{e} and Kranton showed that certain classes of Nash equilibria~\cite{nash50equilibrium}, called \textit{specialized equilibria}, are in one-to-one correspondence with maximal independent sets in the network. These were also shown to be the only `stable' equilibria~\cite{bramoulle2007public}. Recently, Pandit and Kulkarni~\cite{pandit2018refinement} showed an even stronger result  -- that the maximum total effort over all equilibria of such a game was exactly the size of the largest independent set in the network, and they identified independent sets corresponding to `influential' and `frugal' equilibria that well approximated the maximum welfare in such games.
 
A common occurrence in such games, noticed by Bramoull\'{e} and Kranton~\cite{bramoulle2007public}, is the vast multiplicity of equilibria. The number of specialized equilibria is itself exponential in the number of agents or nodes in the network. In addition to specialized equilibria, Bramoull\'{e} and Kranton~\cite{bramoulle2007public} identify \textit{distributed} and \textit{hybrid} equilibria that are also prevalent in such games,  further increasing the number of equilibria. Yet another recurring phenomenon is the dramatic change in the nature equilibria with small changes in the network structure. Addition of edges, for instance, affects the incentives of the agents that are joined by the edge, but this effect propagates through the network, resulting in more far-ranging effects whose total effect is not easy to ascertain. Our motivation in this paper is that these characteristics make it challenging for a principal or a system planner to ascertain outcomes of this interaction and significantly hampers the planner's ability to design interventions that could alter outcomes. For instance, if the actions of the players are investment levels, a system planner may be interested in knowing the total aggregate investment that can result from such interactions. If the actions of the players are the propensity for crime, a planner may be interested in devising structural changes to the network to reduce the total crime levels. Bramoull\'{e} \etal observed in~\cite{bramoulle2014strategic} that the minimum eigenvalue of the graph is a key characteristic that tells us when the equilibrium would be unique. However, despite this result, not much is understood about aggregate play in such interactions.
Our central contributions in this paper are theorems that capture maximum aggregate play and welfare in terms of graph characteristics. This allows maximum aggregate play and welfare (over all equilibria) to be assessed only from the graph structure, \textit{without} enumeration of all equilibria, thereby significantly reducing the complexity of the planner's problem.

We characterize the maximum aggregate play in strategic interactions on networks within a small band for the model discussed in Bramoull\'{e} \etal~\cite{bramoulle2014strategic}. In this model, players are situated on nodes of a network and choose actions to maximize their benefit less the cost of their effort. The benefit a player derives is a function of the sum of its own effort and the total of the effort of his neighbors, where the latter is discounted by a substitutability factor $\delta \in (0,1)$. 
Thus, unit effort by a neighbor $j$ of a player $i$ is substitutable by `$\delta$' units of effort by player $i$, as far as the benefit of player $i$ is concerned. Such a model applies to studying how firms involved in similar kinds of research choose the amount of investment and effort when the interaction structure is given by a network.

We view the case of perfect substitutability ($\delta=1$) as the base case and $\delta \in (0,1)$ as a small perturbation from this case. We show that for $\delta$ greater than an explicit constant $\eta(G)<1$ that depends only on the underlying graph $G$, the maximum aggregate play over all equilibria is within $e^*\alpha(G)$ and $\estar \left(\alpha(G)+ 1 + \frac{1}{\alpha(G)-1}\right)$. Here $\alpha(G)$ is the size of the largest independent of the graph $G$ (called \textit{independence number}), and 
$e^*$ is the level of action at which marginal benefit equals marginal cost. We also identify the structure of the aggregate play maximizing equilibrium. In the special case when a unique maximum independent set exists in the underlying network, we provide the exact value of maximum aggregate play; indeed it is $e^* \alpha(G)$. We also obtain similar bounds on the maximum welfare attained by any equilibrium. When the underlying network is a tree (a graph without cycles), we obtain explicit formulae for the maximum aggregate play and welfare irrespective of the requirement that $\delta \geq \eta(G).$ 
Thus, using our results the planner need only calculate certain underlying graph parameters, such as the size of the maximum independent, size of the largest clique and know the degree distribution of the nodes, and these suffice to estimate maximum aggregate play and maximum welfare. 

These results also show that alterations to the graph structure that do not change these graph parameters have limited effect on the maximum aggregate play and welfare. In particular, notice that deletion of nodes does not increase the independence number, and hence, is not likely to have a significant effect on the maximum aggregate play. However, deletion of edges can increase the independence number, making deletion of edges the more effective strategy at improving aggregate play.


Part of this paper can be viewed as a logical generalization of previous results in~\cite{pandit2018refinement} which consider the case $\delta=1$, and indeed for  $\delta=1$, our results reduce to those in \cite{pandit2018refinement}. However, this generalization is significantly harder to accomplish mathematically. When $\delta<1$ the nature of equilibria is drastically different from that when $\delta=1.$ The source of the hardness is that when $\delta<1$, no canonical classes of equilibria (like specialized equilibria) are known. In particular, maximal independent sets do not, in general, correspond to equilibria of the game. 
We address this challenge by noticing that equilibria of this game correspond to solutions of a suitably defined \textit{linear complementarity problem} (LCP)~\cite{cottle92linear}. In our recent work~\cite{chadha2018independent}, we characterized $\ell_1$ norm maximizing solutions of this linear complementarity problem and identified the existence of $\alpha(G)$ \textit{independent cliques} that support such solutions. Independent cliques are a set of cliques in the graph such that no two nodes from two distinct cliques are adjacent. They can be thought of as a generalization of independent sets and reduce to the latter when the said cliques are single vertices. We show in \cite{chadha2018independent} via a constructive algorithm that such cliques that support an LCP solution necessarily exist in any graph. When translated to the game in consideration here, these correspond to Nash equilibria where agents that make a non-zero effort form a set of independent cliques and in each such clique, all agents have identical actions. Such equilibria, that we call \textit{independent clique equilibria}, maximize aggregate play. When $\delta=1$, these cliques become degenerate singletons and we recover the result of~\cite{pandit2018refinement}. The requirement of $\delta\geq \eta(G)$ is inherited from \cite{chadha2018independent} and is necessary for the existence of such independent cliques.

For the case when the underlying graph is a tree, we provide bounds on the aggregate play and welfare which hold for a larger range of  $\delta$ and depend only on the number of nodes and their degree. We first provide such bounds for the case when the underlying graph is a line network. A {\it line network (chain)} is a network whose nodes can be listed as $1,2,\hdots,n$ such that the links are $\{i,i+1\}$ for $i = 1,2,\hdots,n-1$. Then we consider {\it starlike} trees which are defined as trees with exactly one node of degree strictly greater than 2. Using these as the building blocks, we give tight bounds on maximum aggregate play and welfare for general trees.

\subsection{Organisation of the paper}

In \cref{sec:model}, we provide some definitions from graph theory and provide the model of the strategic interactions game we will work with. We also state a result relating the Nash equilibria of this game to an LCP. In \cref{sec:aggplaywel}, we give bounds on the maximum aggregate play for general networks by defining Independent Clique Equilibria (ICE). We also give approximations of the maximum welfare by relating it to maximum aggregate play and using the bounds derived on it. In \cref{sec:tree}, we consider the case when the underlying network structure is a tree and provide bounds on the maximum aggregate play and welfare independent of the results of \cref{sec:aggplaywel}. We conclude the paper in \cref{sec:concl}.

\section{Model} \label{sec:model}

We consider the model of strategic interaction over networks as discussed in \cite{bramoulle2014strategic}. A network or graph $G = (V,E)$ is defined by a set of nodes $V$ and a set of links  or edges $E \subseteq V \times V$. Two nodes $i,j \in V$ are said to be {\it neighbours} of each other or {\it adjacent} to each other if $(i,j) \in E$. The neighbourhood of a node $i$ in network $G$, denoted by $N_G(i)$, is the set of all nodes which are adjacent to $i$. A path from node $i$ to node $j$ in a network is a sequence of distinct nodes  $v_0\ (=i),v_1,\hdots,v_m\ (=j)$ which begins with $i$ and ends with $j$ such that $(v_{k},v_{k+1})$ is a link for $k=1,\hdots,m$. For a network $G$  on $n$ nodes, let $A \in \{0,1\}^{n\times n}$ denote its adjacency matrix: \ie, 
\begin{equation*}
a_{ij} \coloneqq A(i,j) =  
\begin{cases}
1 , & {\rm  if } \ (i,j) \in E(G) \\
0 , & {\rm  otherwise }
\end{cases}
\end{equation*}

Now we define the concept of an independent set, which is essential to give characterizations of aggregate play maximizing equilibria. An \textit{independent set} of a network is defined as a set of nodes such that no two of them are adjacent.  A $k$\textit{-dominating set} is defined as a set of nodes $S$ such that all nodes which are not in $S$ have at least $k$ neighbours in $S$, \ie   $|N_G(i) \cap S| \geq k \ \forall i \not\in S$. An independent set which is $1$-dominating is called a \textit{maximal independent set}. A \textit{maximum} independent set is an independent set which has the highest cardinality amongst all independent sets. The cardinality of a maximum independent set in $G$ is denoted by $\alpha(G)$ and is also called the independence number of the network $G$. A {\it clique} of a network is defined as a set of nodes such that there exists an edge between any two nodes in this set.
For a vector $x$ indexed by $V(G)$, let $\sigma(x)$ be its support, i.e. 
\[
\sigma(x) \coloneqq \{i \in V(G)| x_i > 0\}.
\]
Let $\bfone_S$ denote the characteristic vector of set $S$, i.e.
\begin{equation*}
\bfone_S(i) \coloneqq (\bfone_S)_i =  
\begin{cases}
1 , & {\rm  if } \ i \in S \\
0 , & {\rm  otherwise }.
\end{cases}
\end{equation*}

We consider the set of agents to be the set of nodes $V$ of a network $G$ with $|V| = n$. Each agent simultaneously chooses a scalar action $x_i \geq 0$. We denote the vector of efforts of all agents as $x = (x_1,x_2,\dots,x_n)$ and by $x_{-i}$ the efforts of all agents except $i$.  The utility of an agent depends on its own action and those of its neighbours. A substitutability factor $\delta$ determines how much the effort of an agent affects the utility of its neighbour. Let the utility of agent $i$ be denoted by $U_i(x_i,x_{-i}; G, \delta)$.   
\begin{equation*}
 U_i(x_i,x_{-i};G,\delta) = b\Bigg(x_i + \delta \sum_{j \in V}a_{ij}x_j\Bigg) - cx_i,
\end{equation*}
where $b(\cdot)$ is a differentiable strictly concave increasing benefit function and $c > 0$ is the marginal cost. 
  Each agent $i$ chooses action $x_i$ to maximize its own utility. We denote the {\it best response} of agent $i$ by $f_i(x_{-i}; G, \delta)$, \ie it represents the action which maximizes $U_i(\cdot,x_{-i}; G, \delta)$ given $x_{-i}, G$ and $\delta$. It is easy to see that 
\begin{equation}\label{eqn:bestreply}
f_i(x_{-i}; G, \delta) =  
\begin{cases}
\estar  - \delta \sum_{j \in V}a_{ij}x_j , & {\rm  if } \ \delta \sum_{j \in V}a_{ij}x_j < \estar , \\
0 , & {\rm  if } \ \delta \sum_{j \in V}a_{ij}x_j \geq \estar, \\
\end{cases}
\end{equation}
where $\estar$ is such that the marginal benefit is equal to the marginal cost at $\estar$, \ie $b'(\estar) = c$. The best response can be succinctly written as
\begin{equation}\label{eqn:bestreply2}
f_i(x_{-i}; G, \delta) =  
\max\{0, \estar  - \delta \sum_{j \in V}a_{ij}x_j\}
\end{equation}
Our results apply to all games where the best response is of the above `max-linear' form. Thus, an agent has an incentive to make positive effort if and only if the sum of the efforts of his neighbors is strictly less than $e^*/\delta;$ else the agent makes $0$ effort. 

\subsection{Characterization of Nash equilibria of the game}
A Nash equilibrium is a profile of efforts $x^*=(x_1^*,\hdots,x_n^*)$ for which $x_i^* = f_i(x^*_{-i}; G, \delta)$ for all agents $i \in V$. 
The set of Nash equilibria of the game is denoted by $\NE_{\delta}(G)$. 
We now provide a characterization of the Nash equilibria of the game using a linear complementarity problem. 
To this end, we define 
$$\C_i(x) \coloneqq x_i + \delta \sum_{j \in V}a_{ij}x_j = x_i + \delta \sum_{j \in N_G(i)}x_j$$ and call it the {\it discounted sum of the closed neighbourhood} of $i$ with respect to $x$. We also call $\C_i(x) - x_i$ as the {\it discounted sum of open neighbourhood} of $i$ with respect to $x$. Let $\C(x) $ denote the column vector $ (\C_1(x),\C_2(x),\dots,\C_n(x))$. Then, $\C(x) = (I + \delta A)x$ where $I$ denotes the $n \times n$ identity matrix and $A$ denotes the adjacency matrix of the network $G$. $x$ is an equilibrium of the above game if $x_i = f_i(x_{-i}; G, \delta) \ \forall i \in V$. From \eqref{eqn:bestreply2} it is evident that $x$ is a Nash equilibrium of the game if and only if $x$ satisfies
\begin{enumerate}
 \item $x_i  = 0$ and $\C_i(x) \geq \estar$.
 \item $x_i  > 0$ and $\C_i(x) = \estar$
\end{enumerate}
These conditions can be equivalently written as 
\[
x_i \geq 0, \  \C_i(x) \geq \estar, \  x_i(\C_i(x) - \estar) = 0 \ \ \forall i \in V .
\] 
In the vector form these conditions are equivalent to 
\begin{equation}\label{eqn:eqmlcp}
x \geq 0, \ \C(x) \geq \estar\bfe, \ x \t (\C(x) - \estar\bfe) = 0,  
\end{equation} 
where $\bfe=(1,1,\dots,1)$ is the vector of 1's. 

Conditions in \eqref{eqn:eqmlcp} correspond to that of a linear complementarity problem. Given any matrix $M \in \Real^{n \times n}$ and a vector $q \in \Real^n$, LCP($M,q$) is the following problem:
\[
 {\rm Find } \ x \in \Real^n \  {\rm such} \ {\rm that} \ \ x \geq 0, \ y = Mx + q \geq 0, \ y\t x =  0.
\]
The vector $x$ is called a solution of this problem. 
Let SOL(LCP($M,q$)) denote the set of $x$ such that $x$ solves the LCP($M,q$). Specifically, for $M = I+\delta A$ and $q = -\bfe$, $\LCP(I+\delta A,-\bfe)$ is given by
\[
{\rm Find } \ x \in \Real^n \  {\rm such} \ {\rm that} \ \ x \geq 0, \ (I + \delta A)x \geq \bfe, \ ((I + \delta A)x - \bfe)\t x =  0.
\]
We denote $\LCP(I+\delta A,-\bfe)$ by $\LCP_\delta(G)$ and the  set of its solutions by $\SOL_\delta(G)$. The relation between Nash equilibria of the game and the solutions of $\LCP_\delta(G)$ are given by the following theorem, which  follows immediately from \cref{eqn:eqmlcp}.

\begin{theorem}\label{thm:nashlcp}
An action profile $x$ is a Nash equilibrium of the game if and only if 
\[
\frac{1}{\estar}x \in \SOL_\delta(G). 
\]
\end{theorem}

The authors in \cite{bramoulle2007public} considered a special case of the game with $\delta = 1$. They found a class of Nash equilibria, called specialized equilibria, in which each agent exerts either the maximum effort $\estar$ (called specialists) or no effort (such agents were called free-riders). They proved that a profile is a specialized Nash equilibrium if and only if the specialists form a maximal independent set of the underlying network. We note that such a result is not true for the more general game we consider. In conjunction with \cref{thm:nashlcp} and \cite[Lemma 3.2]{chadha2018independent}, it can be shown that a profile is a specialized Nash equilibrium (as defined in \cite{bramoulle2007public}), if and only if the specialists form a $\lceil \frac{1}{\delta} \rceil$-dominating independent set of the underlying network. A network may not have a $\lceil \frac{1}{\delta}\rceil$-dominating independent set, in which no specialized equilibrium can exist.

\subsection{The game with $\delta = 1$}

When we take the substitutability factor $\delta $ as unity, the model reduces to the game of provision of non-excludable public goods in the presence of a network as considered in \cite{bramoulle2007public}. We review some results related to this model here.

Define the concavity of the benefit function as
\begin{equation}\label{eqn:ccvdefpublic}
 \rho = \frac{b(n\estar) - b(\estar)}{c(n-1)\estar}.
\end{equation}
The welfare is defined as
\[
 W(x;G) = \sum_{i \in V}b\biggl(x_i + \sum_{j \in N_G(i)}x_j\biggr) - c\sum_{i \in V}x_i
\]
for an effort profile $x$ when the underlying network is $G$. Bramoull\'{e} and Kranton prove the following proposition on comparing the welfare of two effort profiles under suitable conditions on $\rho$.

\begin{proposition}
 Consider two effort profiles $x^1$ and $x^2$ on G. Then there exists $\rho' < 1$ such that for fixed $b(\estar)$ and $b'(\estar)$, and $\rho > \rho'$, we have $W(x^1;G) > W(x^2;G)$ if $\sum_{i \in V}d_ix^1_i > \sum_{i \in V}d_ix^2_i$, where $d_i$ is the degree of node $i$.
\end{proposition}

The authors of \cite{pandit2018refinement} took this work further and proved that the effort maximizing equilibrium among all equilibria is a specialized equilibrium. 
They proved that the maximum weighted effort amongst all equilibria was attained by a specialized equilibrium. Here, the weighted  effort of effort profile $x \in \Real^n$ is defined as $\sum_{i \in V}w_ix_i$, where $w = (w_1,w_2,\dots,w_{n}) > 0$ denotes the set of weights. 
We state some of the results of \cite{pandit2018refinement} below.

The first theorem we recall from \cite{pandit2018refinement} proves that the maximum weighted effort in a public goods game is achieved by a specialized equilibrium and characterizes that specialized equilibrium which achieves the maximum. It uses the notion of a $w$-weighted maximum independent set. Given a set of weights $w = (w_1,w_2,\dots,w_{|V})$, the weight of an independent set $S \subseteq V$ is defined as $\sum_{i \in S}w_i$.  An independent with the maximum weight is called a $w$-weighted maximum independent set and the maximum value is called the $w$-weighted independence number, denoted by $\alpha_w(G)$.

\begin{theorem}\label{thm:partheeffortmax}
 Consider a public goods game over a network $G$, with $w = (w_1,w_2,\dots,w_{|V})$ denoting the set of weights and $w_i > 0 \ \forall i \in V$. If $S^*$ is a $w$-weighted maximum independent set, then the specialized equilibrium in which all agents in $S^*$ are specialists and others free ride is an effort maximizing equilibrium.
\end{theorem}

The authors of \cite{pandit2018refinement} also proved that the maximum weighted welfare in a public goods game is achieved by a specialized equilibrium in a limiting case as the concavity of the benefit function, as defined in \cref{eqn:ccvdefpublic}, tends to 1. They also provide expressions for the maximum weighted welfare in terms of the weighted independence number of the network for the same. The weighted welfare in a public goods game is defined as $W_w(x;G) = \sum_{i \in V}w_ib(x_i + \sum_{j \in N_G(i)}x_j) - c\sum_{i \in V}w_ix_i$, where $w = (w_1,w_2,\dots,w_{|V|})$ denotes the set of weights.

\begin{theorem}
 Consider a public goods game over a network $G$ without isolated agents and $w = (w_1,w_2,\dots,w_{|V})$ denoting the set of weights with $w_i > 0 \ \forall i \in V$. If $S^*$ is a $Aw$-weighted maximum independent set, where $A$ is the adjacency matrix of the network $G$ and let $x^*$ be the specialized equilibrium in which all agents in $S^*$ are specialists and others free ride. Then, as $\rho \rightarrow 1$, the welfare of $x^*$ approaches the maximum equilibrium welfare.
 \[
  \lim_{\rho \rightarrow 1}W^*_w = \lim_{\rho \rightarrow 1}W_w(x^*) = n(b(\estar) - c\estar) + c\estar\alpha_{Aw}. 
 \]
\end{theorem}
\noindent The specialized equilibrium corresponding to the $Aw$-weighted maximum independent set was referred to as the \textit{influential equilibrium}. The authors in~\cite{pandit2018refinement} also consider the case where $\rho \rightarrow 0$, in which another specialized equilibrium, called the \textit{frugal} equilibrium gives the maximum welfare.

\subsection{Results on $\LCP_\delta(G)$}\label{sec:lcpresults}

In this section, we prove some results on the solutions of LCP$_\delta(G)$ and we recall some results from \cite{chadha2018independent} on the solutions of $\LCP_\delta(G)$ which we use extensively in our analysis. We first prove that if we restrict the network to all nodes except one which contributed $0$ in a solution, then the restricted vector is a solution of the $\LCP_{\delta}(G')$, where $G'$ is the restricted network. We also prove that if a network is a disjoint union of connected components~\cite{west00introduction},  \cite{bondy1976graph}, any solution restricted to a component is a solution of the LCP on the network being that component.

\begin{lemma}\label{lem:lcp_genprop}
Consider $\LCP_{\delta}(G)$ and let $x \in \SOL_\delta(G)$. Then,
\begin{enumerate}[label = \alph*.]
\item If $x_i = 0$ for some $i \in V(G)$, then $x_{-i} \in \SOL_{\delta}(G')$, where $G' = G_{-i}$ \ie network $G$ restricted to all nodes except $i$.
\item If $G = \cup_{i = 1}^{k}G_i$, where $G_1,\hdots, G_k$ are connected components of $G$, then $x_{G_i} \in \SOL_\delta(G_i)$ for $i=1,\hdots,k$.
\end{enumerate}
\end{lemma}
\begin{proof}
We know that $x$ satisfies $x_j \geq 0, x_j + \delta \sum_{m \in N_G(j)}x_m \geq 1$ and $x_j(x_j + \delta \sum_{m \in N_G(j)}x_m - 1) = 0 \ \forall j \in V(G)$.
\begin{enumerate}[label = \alph*.]
\item  Since $x_i = 0$, we have $\sum_{m \in N_G(j)}x_m = \sum_{m \in N_{G'}(j)}x_m \ \forall j \in V(G')$. Thus, for $y = x_{-i}$ we have $y_j \geq 0, y_j + \delta \sum_{m \in N_{G'}(j)}y_m \geq 1$ and $y_j(y_j + \delta \sum_{m \in N_{G'}(j)}y_m - 1) = 0 \ \forall j \in V(G')$ giving $x_{-i} \in \SOL_{\delta}(G')$.
\item We note that  $\sum_{m \in N_G(j)}x_m = \sum_{m \in N_{G_i}(j)}x_m \ \forall j \in V(G_i) \ \forall i \in \{1,2,\dots,k\}$. This is because $N_G(j) = N_{G_i}(j) \ \forall j \in V(G_i)$ since all nodes in a component have neighbours only in that component. Thus,  for $y = x_{G_i}$ we have $y_j \geq 0, y_j + \delta \sum_{m \in N_{G_i}(j)}y_m \geq 1$ and $y_j(y_j + \delta \sum_{m \in N_{G_i}(j)}y_m - 1) = 0 \ \forall j \in V(G_i) \ \forall i  \in \{1,2,\dots,k\}$ giving $x_{G_i} \in \SOL_{\delta}(G_i) \ \forall i \in \{1,2,\dots,k\}$.
\end{enumerate}
\end{proof}

We now define Independent Clique Solutions (ICS), a concept introduced in \cite{chadha2018independent} which forms the basis of our analysis.

\begin{definition}
An \textit{Independent Clique Solution (ICS)} is a vector in $\Real^{|V|}$ which is a solution of the LCP$_\delta(G)$ for some $\delta < 1$ and its support is a union of independent cliques, where two cliques in a network are said to be \textit{independent} if no node of one clique has any node of the other clique as its neighbour.
\end{definition}

The set of all independent clique solutions of $\LCP_\delta(G)$ is denoted by ICS$_{\delta}(G)$. An algorithm to find these solutions has been given in \cite{chadha2018independent}. The input to this algorithm is the network which underlies the game, a maximum independent set of the network and the substitutability factor $\delta$. Since a maximum independent set always exists for any graph, ICSs of $\LCP_\delta(G)$ exist for any $G$ and suitable $\delta$. The algorithm iterates over nodes of the given maximum independent set and for each node (say $i$), the solution support is updated from that node to a clique involving $i$'s neighbours which have only $i$ as a neighbour in the maximum independent set. The results of \cite{chadha2018independent} dictate that if the support of an ICS is the set of independent cliques given by $\Kscr = \{C_1, C_2,\hdots, C_{|\Kscr|}\}$, then the ICS supported on $\Kscr$, if it exists, is given by
 \begin{equation*}\label{eqn:icsfundef}
   x_i = \begin{cases}
                    \frac{1}{1 + (|C_j|-1)\delta}, \ & \ {\rm if} \ i \in C_j, \ C_j \in \mathcal{K}\\
                    0, \ & \ {\rm otherwise}.
                   \end{cases}  
 \end{equation*}

The authors of \cite{chadha2018independent} have proved that the solution which maximizes the $\ell_1$ amongst the independent clique solutions  also maximizes the $\ell_1$ amongst all solutions. The $\ell_1$ norm of a vector $x$ is defined as the sum of all elements of $x$, denoted by $||x||_1$, \ie $||x||_1 = \sum_{i \in V}x_i$. These results hold in a certain range of $\delta$ which is defined using $\eta(G)$ given below
\begin{eqnarray}\label{eqn:etadef}
 \eta(G) =  \max\Big{\{}\frac{\omega(G)-3 + \sqrt{(\omega(G)-3)^2 + 4(\omega(G)-1)}}{2(\omega(G)-1)},\\ 
 \nonumber \frac{\alpha(G)(\omega(G)-1) - \omega(G)}{\alpha(G)(\omega(G)-1)}\Big{\}}.
\end{eqnarray}

\begin{theorem}\label{thm:maxicsmor}
Let  $\delta \in [\eta(G),1]$, and let $x^* \in$ ICS$_\delta(G)$ be an ICS that maximizes $\sum_{i \in V}x_i$ amongst all $x \in {\rm ICS}_\delta(G)$. Then $x^*$ also maximizes  $\sum_{i \in V}y_i$ amongst all $y \in {\rm SOL}_\delta(G)$.
\end{theorem}

We also use a corollary of the above theorem for the special case of networks in which unique maximum independent sets exist. In such a case, it is proved that the characteristic vector of the unique maximum independent set maximizes the $\ell_1$ norm amongst all solutions.

\begin{corollary}\label{cor:uniqindmor}
 Let $G$ be a graph with a unique maximum independent set $S$. Then, for $\delta \in [\eta(G),1]$, $1_S$ maximizes the $\ell_1$ norm amongst all solutions of $\LCP_\delta(G)$.  
\end{corollary}
\noindent With this we conclude our background. We now proceed to the main results of the paper.

\section{Aggregate play and welfare}\label{sec:aggplaywel}

In this section, we first find tight lower and upper bounds on the maximum aggregate play in terms of the properties of the network (\cref{sec:aggplay}). Then we use these bounds to approximate the maximum welfare (\cref{sec:aggwel}). 

\subsection{Maximum aggregate play} \label{sec:aggplay}
In this section, we bound the maximum aggregate play attainable in any equilibrium within limits that depend on $\alpha(G).$ In addition, we identify the structure of an aggregate play maximizing equilibrium. We first discuss a lower bound on the maximum aggregate play which holds for all networks and all values of $\delta$. Then, we give a general form of the aggregate play maximizing solution for any network under suitable conditions on $\delta$. For this, we use the results of \cite{chadha2018independent}, wherein the concept of independent clique solutions (ICS), which are a special class of solutions of $\LCP_\delta(G)$ was introduced. Through this, we get an absolute upper bound on the maximum aggregate play depending on the independence number of the network. 

The aggregate effort under a strategy profile $x$ is defined as the sum over efforts of all nodes of the given graph $G$, \ie
\[
 E(x;G) \coloneqq  \sum_{i \in V}x_i
\]
 Let $E^*(\delta,G)$ denote the maximum aggregate play over all equilibria. We first ask, how large can $E^*(\delta,G)$ be? In particular, does the maximum aggregate play display any monotonicity with $\delta$? We find that with decrease in $\delta$ the maximum aggregate play \textit{increases}. As $\delta$ decreases the efforts of neighbors contribute lesser to the benefit of a player, forcing the player to compensate with higher effort himself. As a result, in equilibrium,  players on the whole make higher effort. This was shown in \cite[Prop 8]{bramoulle2014strategic}.
\begin{theorem}\label{thm:decrplay}
Let $E^*(\delta,G)$ be the maximium aggregate play of the game with parameter $\delta$ and network $G$. If $\delta' \geq \delta$,
\[
E^*(\delta',G) \leq E^*(\delta,G).
\] 
\end{theorem}
Combining this result with \cref{thm:partheeffortmax}, we get an an absolute lower bound on the maximum aggregate effort.

\begin{theorem}\label{thm:lower}
Consider the game with substitutability parameter $\delta < 1$ and network $G$. Then we have,
\[
E^*(\delta,G) \geq \estar \alpha(G),
\]
where $\alpha(G)$ is the independence number of the graph.
\end{theorem}
\begin{proof}
From \cref{thm:decrplay}, the maximum aggregate play for any game with $\delta \leq 1$ will be at least as much as the maximum aggregate play in the case when $\delta = 1$. For $\delta = 1$, we know from \cref{thm:partheeffortmax} that  the maximum aggregate play is $\estar \alpha(G)$. Thus, the maximum aggregate play for $\delta \leq 1$ at least $\estar\alpha(G)$.
\end{proof}

\subsubsection{Independent Clique Equilibria (ICE)}

We have seen that the aggregate play increases with a decrease in $\delta$. We now study how aggregate play maximizing equilibria change as $\delta$ decreases below unity.  
It is instructive to begin this discussion with the simplest case -- when the underlying network is a complete graph (every pair of nodes has an edge between them). 

Consider a complete graph $G$ with $n$ nodes and with $\delta = 1$. For this case, the specialized equilibrium in which any one agent, say agent $1$, plays $e^*$ and others free-ride is an equilibrium. Indeed, this also is an aggregate play maximizing equilibrium yielding aggregate play $e^*$. There is also a distributed equilibrium in which all agents make effort to $\frac{e^*}{n}$, and this also yields the same aggregate play. Now consider $\delta<1$. In this case, the agents other than agent $1$, who were free-riders when $\delta$ was 1, cannot continue to free-ride, since LCP conditions \cref{eqn:eqmlcp} do not hold. In fact, no equilibrium exists in which free-riding is allowed. Consequently, erstwhile free-riders must now exert positive effort. This, in turn, means that the erstwhile specialist agent $1$ can \textit{reduce} its effort, and makes effort strictly less than $e^*$. The aggregate play maximizing equilibrium now involves all agents making effort equal to $\frac{\estar}{1 + (n-1)\delta}$.  The resulting maximum aggregate play is $\frac{n\estar}{1 + (n-1)\delta}$, which for $\delta<1$ is strictly larger than the maximum aggregate play for $\delta=1.$

\begin{figure}
 \centering
\begin{tikzpicture}[auto, node distance=2cm, every loop/.style={},
                    thick,main node/.style={circle,draw,font=\sffamily\small\bfseries},dark node/.style={circle,draw,fill = black!40,font=\sffamily\small\bfseries}]
 
   \node[main node] (1) {$1$};
  \node[main node] (2) [above right of=1]{$2$};
  \node[main node] (3) [below right of=2] {$3$};
  \node[main node] (4) [below left of=3] {$4$};

  \path[every node/.style={font=\sffamily\small}]
    (1) edge node [right] {} (2)
    (2) edge node [right] {} (3)
    (3) edge node [right] {} (4)
    (4) edge node [right] {} (1)
    (4) edge node [right] {} (2)
    (1) edge node [right] {} (3);  
\end{tikzpicture}
\caption{Example of a complete support equilibrium.}
\label{fig:ics_completeex}
\end{figure}
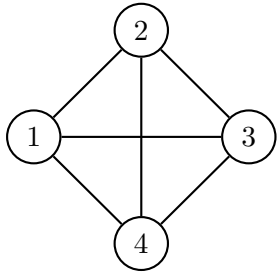

\begin{examplec}\label{ex:cliq}
\cref{fig:ics_completeex} shows a complete network with 4 nodes. Consider a maximum independent set of the graph in \cref{fig:ics_completeex}, say $S = \{1\}$. Then, $ \estar \bfone_{S}= \left( \estar,0,0,0\right)$ is an equilibrium for the game with $\delta = 1$.

When $\delta < 1$, $ \estar \bfone_{S}$ is no longer an equilibrium of the game. This is because, for every node other than $\{1\}$, the net play observed is $\delta \estar$  and since $\delta \estar < \estar$ the equilibrium conditions are not satisfied. Thus nodes other than node $1$ have to contribute positively. This provides incentive to node $\{1\}$ to contribute lesser than $\estar$. To formally prove that each node contributes a positive amount in any equilibrium, assume any one node (say $4$) to contribute $0$ and others to contribute $x_1,x_2,x_3$ respectively. Then, using the equilibrium conditions on $4$, we have 
\begin{equation}
\delta(x_1 + x_2 + x_3) \geq \estar. \label{eq:ex1} 
\end{equation}
Without loss of generality, we assume $x_1$ is non-zero since all cannot play 0.
Now, the equilibrium conditions on $1$ dictate $\C_1(x) = \estar$, \ie $\estar = x_1 + \delta(x_2 + x_3)$. Thus using \eqref{eq:ex1}, we get $\estar \geq \estar + x_1(1 - \delta)$, which is not possible. Thus, each node contributes a positive amount. It is then easy to see that the only equilibrium is one in which each node contributes equally $x = \left( \frac{\estar}{1 + 3\delta}, \frac{\estar}{1 + 3\delta}, \frac{\estar}{1 + 3\delta}, \frac{\estar}{1 + 3\delta}\right)$. The aggregate play is now $\frac{4\estar}{1+3\delta}> \estar.$
\end{examplec}

We find that this phenomenon occurs more generally in all networks. When $\delta=1$, the specialized equilibrium supported by a maximum independent set maximizes aggregate play (thanks to Theorem~\ref{thm:partheeffortmax}). When $\delta<1$, certain neighbors of specialists in this equilibrium, who where free-riders for $\delta=1$, exert positive effort and all specialists in turn reduce their effort. For $\delta<1$ but close to unity, the aggregate play maximizing equilibrium is supported on $\alpha(G)$-many cliques, each containing one node from the aforesaid maximum independent set. We illustrate this through the example below.

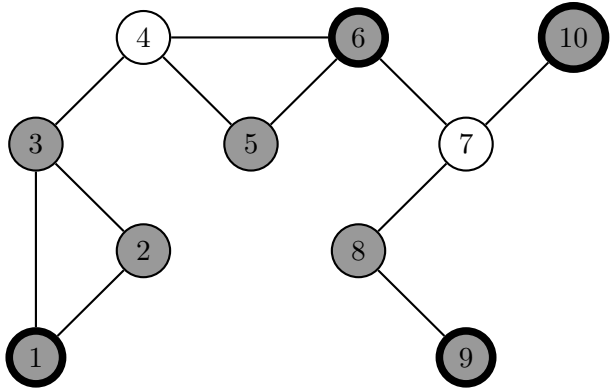
\begin{figure}
 \centering
\begin{tikzpicture}[auto, node distance=2cm, every loop/.style={},
                    thick,main node/.style={circle,draw,font=\sffamily\small\bfseries},dark node/.style={circle,draw,fill = black!40,font=\sffamily\small\bfseries}, ind node/.style={circle,draw,fill = black!40 ,font=\sffamily\small\bfseries, line width=1mm}]
 
   \node[ind node] (1) {$1$};
  \node[dark node] (2) [above right of=1]{$2$};
  \node[dark node] (3) [above left of=2] {$3$};
  \node[main node] (4) [above right of=3] {$4$};
  \node[dark node] (5) [below right of=4] {$5$};
  \node[ind node] (6) [above right of=5] {$6$};
  \node[main node] (7) [below right of=6] {$7$};
  \node[dark node] (8) [below left of=7] {$8$};
  \node[ind node] (9) [below right of=8] {$9$};
  \node[ind node] (10) [above right of= 7] {$10$};

  \path[every node/.style={font=\sffamily\small}]
    (1) edge node [right] {} (2)
    (2) edge node [right] {} (3)
    (3) edge node [right] {} (4)
    (4) edge node [right] {} (5)
    (4) edge node [right] {} (6)
    (5) edge node [right] {} (6)
    (6) edge node [right] {} (7)
    (7) edge node [right] {} (8)
    (9) edge node [right] {} (8)
    (10) edge node [right] {} (7)
    (3) edge node [right] {} (1);
\end{tikzpicture}
\caption{Example of an independent clique equilibrium (ICE). The support is represented by dark nodes and the nodes with a bold outline denote a related maximum independent set which is the support of an example solution when $\delta = 1$.}
\label{fig:maxics_ex}
\end{figure}

\begin{examplec}\label{ex:ice}
Consider the network in \cref{fig:maxics_ex} and the maximal independent set formed by the highlighted nodes, $S = \{1,6,9,10\}$. Then, $\estar \bfone_{S}= \left( \estar,0,0,0,0,\estar,0,0,\estar,\estar \right)$ is an equilibrium for the game with $\delta = 1$.

When $\delta < 1$, $ \estar \bfone_{S}$ is no longer an equilibrium of the game. This is because, for nodes $2,3,4,5$ and $8$ the net effort is $\delta \estar$  and since $\delta \estar < \estar$ the equilibrium conditions are not satisfied. We notice that one  equilibrium is given by $x = \Big( \frac{\estar}{1 + 2\delta},\frac{\estar}{1 + 2\delta},\frac{\estar}{1 + 2\delta},0,$ $\frac{\estar}{1 + \delta},\frac{\estar}{1 + \delta},0,\frac{\estar}{1 + \delta},\frac{\estar}{1 + \delta},\estar \Big)$. 

Since the all the nodes connecting the network restricted to $\{1,2,3\}$ to the other parts of the network contribute $0$, we can isolate the nodes $\{1,2,3\}$ and the corresponding restricted network and analyze its equilibrium conditions. Note that amongst $\{1,2,3\}$, $\{2\}$ and $\{3\}$ have only one neighbour in the independent set $S$. Thus, for $\delta < 1$ with only $x_1 = \estar$ and $x_2 = x_3 = 0$, equilibrium conditions are not satisfied. This provides them incentive to contribute positively. Since this is true for every node in the independent set, the support of the solution blows up from an independent set to set of independent cliques. 
\end{examplec}

We formalize this notion of blowing up of independent nodes into cliques by defining a special class of equilibria which we call independent clique equilibria (ICE). The definition of ICE is based on the independent clique solutions (ICS), defined in \cref{sec:lcpresults}.

\begin{definition}
An \textit{Independent Clique Equilibrium (ICE)} is an action profile $x \in \Real^{|V|}$ such that $\frac{1}{\estar}x$ is an ICS of $\LCP_\delta(G)$.
\end{definition}
Thus, given the support as a union of independent cliques $\Kscr = \{C_1,C_2,$ $\hdots,C_{|\Kscr|}\}$, the only possible ICE is given by 
\begin{equation*}\label{eqn:icsfundef}
   x_i = \begin{cases}
                    \frac{1}{1 + (|C|-1)\delta}\estar, \ & \ {\rm if} \ i \in C, \ C \in \mathcal{K}\\
                    0, \ & \ {\rm otherwise}.
                   \end{cases}  
 \end{equation*}
 The class of ICEs with $\alpha(G)$ cliques is aggregate play maximizing. 

\begin{theorem}\label{thm:maxics}
For any network $G$, if $\delta \geq \eta(G)$, then the maximum aggregate play amongst the Nash equilibria of the game is achieved by an independent clique equilibrium with $\alpha(G)$ cliques, where $\eta(G)$ is as defined in \cref{eqn:etadef}.
\end{theorem}
\begin{proof}
From \cref{thm:nashlcp}, we know that $x$ is a Nash equilibrium of the game if and only if $\frac{1}{\estar}x$ is a solution of $\LCP_\delta(G)$.  From \cref{thm:maxicsmor}, we know that the $\ell_1$ norm maximizing solution of $\LCP_\delta(G)$ is an independent clique solution (say ${y}$). Hence, the aggregate play maximizing Nash equilibrium of the game is $\estar{y}$, an ICE.
\end{proof}

\cref{thm:maxics} shows that the maximum aggregate play is achieved by an ICE. We have also seen from \cref{thm:lower} that the maximum aggregate play is at least $\estar \alpha(G)$. The following theorem shows that it is not much larger for $\delta\geq \eta(G)$. 

\begin{theorem}\label{thm:absmax}
Consider the game with substitutability parameter $\delta \in [\eta(G),1]$ and network $G$. Then we have,
\[
\estar \alpha(G) \leq E^*(\delta,G) \leq \estar\left( \alpha(G) + 1 + \frac{1}{\alpha(G)-1}\right)
\]
where $\alpha(G)$ is the independence number of the graph.
\end{theorem}

We now consider the special case of networks for which a unique maximum independent set exists. In this case, the cliques that specialist nodes blowup into must contain only the specialist node. If not, by choosing distinct nodes from any one clique would yield distinct maximum independent sets, contradicting uniqueness. In this case, under suitable conditions on $\delta$, the characteristic vector of the unique maximum independent set is an equilibrium and it is aggregate play maximizing. Moreover, the aggregate play equals $\estar \alpha(G).$ 

\begin{theorem}\label{thm:uniqmax}
 If $G$ has a unique maximum independent set ($S$), for $\delta \geq \eta(G)$ then the maximum aggregate play amongst the Nash equilibria of the game is $\estar\alpha(G)$ and is achieved by the characteristic vector of $S$.
\end{theorem}
\begin{proof}
From \cref{thm:nashlcp}, we know that $x$ is a Nash equilibrium of the game if and only if $\frac{1}{\estar}x$ is a solution of $\LCP_\delta(G)$.  From \cref{cor:uniqindmor}, we know that the $\ell_1$ norm maximizing solution of $\LCP_\delta(G)$ is the characteristic vector of the unique maximum independent set ($\bfone_S$). Hence, the aggregate play maximizing Nash equilibrium of the game is $\estar\bfone_S$ and the maximum aggregate play is given by $\estar\alpha(G)$.
\end{proof}

\subsection{Maximum welfare} \label{sec:aggwel}

We now derive approximations to the maximum welfare. We consider only unweighted welfare. The welfare is defined as
\[
 W(x;\delta,G) = \sum_{i \in V}b\biggl(x_i + \delta\sum_{j \in N_G(i)}x_j\biggr) - c\sum_{i \in V}x_i
\]
for an effort profile $x$ when the underlying network is $G$.  Let $W^*(\delta,G)$ denote the maximum welfare over all equilibria. 

\begin{figure}
\centering
\includegraphics[scale = 0.5]{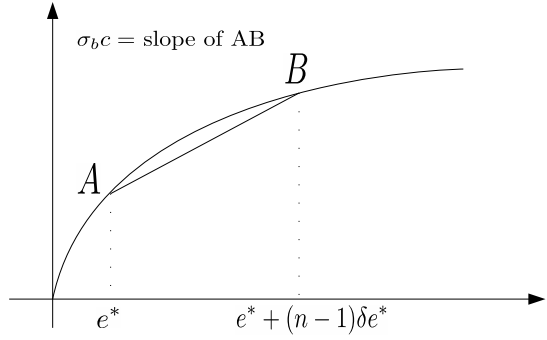}
\caption{A figure illustrating the definition of concavity.}\label{fig:sigmadef}
\end{figure}

To relate the welfare to aggregate play, we define the concavity $\sigma_b$ of the payoff function similar to that defined in \cite{bramoulle2007public} as follows.
\begin{equation}\label{eqn:concavity_def}
 \sigma_b \coloneqq \frac{b(\estar + \delta(n-1)\estar) - b(\estar)}{c(n-1)\estar\delta},
\end{equation}
where $n$ is the number of agents. It is the slope of the secant between $e^*$ and $\estar + \delta(n-1)\estar$ normalized by $c$ so that $\sigma_b < 1$ (\cref{fig:sigmadef}). Note that since $b(\cdot)$ is increasing, we have $\sigma_b > 0$. By strict concavity of $b(\cdot)$, we have $\frac{b(\estar + \delta(n-1)\estar) - b(\estar)}{\delta(n-1)\estar} < b'(\estar) = c$, where the last equality follows from the definition of $\estar$. Thus, we have $0 <\sigma_b < 1$.

Our aim is to find approximations of welfare in terms of aggregate play and then bound the welfare using bounds on aggregate play. The following inequality is central to our analysis:
\begin{equation}\label{eqn:linapprox}
 b(\estar) + c\sigma_b(\C_i(x) - \estar) \leq b(\C_i(x)) \leq b(\estar) + c(\C_i(x) - \estar),
\end{equation}
for all $i \in V$, for any $x$ that is an equilibrium of the game. To prove the first inequality in \cref{eqn:linapprox}, we note that since $x$ is an equilibrium, $\C_i(x) \leq \estar + \delta(n-1)\estar$. Since $b(\cdot)$ is a concave function, we have 
\[
\frac{b(\estar + \delta(n-1)\estar) - b(\estar)}{(n-1)\estar\delta} \leq \frac{b(\C_i(x)) - b(\estar)}{\C_i(x) - \estar},
\]
which implies $b(\estar) + c\sigma_b(\C_i(x) - \estar) \leq b(\C_i(x))$. Also, we know that $c = b'(\estar)$ is the slope of the tangent at $\estar$ and since $b(\cdot)$ is concave, the tangent lies above the secant giving, 
\[
 \frac{b(\C_i(x)) - b(\estar)}{\C_i(x) - \estar} < b'(\estar) = c,
\]
which implies $b(\C_i(x)) \leq b(\estar) + c(\C_i(x) - \estar)$. Thus, we have proved inequalities stated in \cref{eqn:linapprox}. Taking limits as $\sigma_b$ tends to 1 in \cref{eqn:linapprox}, fixing $\estar$ and $b'(\estar)$ we get
\begin{equation}
 \lim_{\sigma_b \rightarrow 1} b(\C_i(x)) = b(\estar) + c(\C_i(x) - \estar).
\end{equation}

We now find approximations of the total welfare for a given Nash equilibrium using expressions involving the total effort. 

\begin{lemma}\label{lem:welbound}
Consider a network $G$ with minimum and maximum degree $d_{min}$ and $d_{max}$ respectively and substitutability factor $\delta \geq \eta(G)$. Let $x$ be a Nash equilibrium for such a game. Then, we have 
 \begin{equation}\label{eqn:welbound1}
  n(b(\estar) - c\estar) + c((d_{min}\delta + 1)\sigma_b - 1)\sum_{i \in V} x_i \leq W(x;\delta,G),
  \end{equation}
  \begin{equation} \label{eqn:welbound2} 
   W(x;\delta,G) \leq  n(b(\estar) - c\estar) + cd_{max}\delta\sigma_b\sum_{i \in V} x_i.
 \end{equation}
\end{lemma}

The above lemma shows that the welfare of any equilibrium can be bounded within two linear functions of the aggregate effort of the equilibrium. 
Using \cref{lem:welbound} and \cref{thm:absmax}, we find bounds on the maximum welfare.

\begin{theorem}\label{thm:welapprox}
Consider a strategic interactions game on a network $G$ with minimum and maximum degree  $d_{min}$ and $d_{max}$ respectively and substitutability factor $\delta \geq \eta(G)$. Then, we have the following:
\begin{enumerate}[label = (\alph*)]
\item For $\sigma_b \in (0,1)$, we have
\begin{equation}\label{eqn:maxuppbnd}
 W^*(\delta,G) \leq n(b(\estar) - c\estar) + cd_{max}\delta\sigma_b\estar\left(\alpha(G)+1 + \frac{1}{\alpha(G)-1}\right).
\end{equation}
 \item  For $\sigma_b \leq \frac{1}{1+d_{min}\delta}$, we have
\begin{equation}\label{eqn:maxlowbnd1}
  W^*(\delta,G) \geq n(b(\estar) - c\estar) + c((d_{min}\delta + 1)\sigma_b - 1)\estar\left(\alpha(G)+1 + \frac{1}{\alpha(G)-1}\right).
\end{equation}
\item  For $\sigma_b \geq \frac{1}{1+d_{min}\delta}$, we have
\begin{equation}\label{eqn:maxlowbnd2}
  W^*(\delta,G) \geq n(b(\estar) - c\estar) + c((d_{min}\delta + 1)\sigma_b - 1)\estar\alpha(G).
\end{equation}
\item If $G$ has a unique maximum independent set, we have 
\begin{equation*}
 W^*(\delta,G) \leq n(b(\estar) - c\estar) + cd_{max}\delta\sigma_b\estar \alpha(G)
\end{equation*}
and 
\begin{equation*}
  W^*(\delta,G) \geq n(b(\estar) - c\estar) + c((d_{min}\delta + 1)\sigma_b - 1)\estar \alpha(G),
\end{equation*}
$ \forall \sigma_b \in (0,1)$.
\item If $G$ is a regular graph, \ie $d_{max} = d_{min} = d$, then keeping $b(\estar)$ and $b'(\estar)$ fixed and letting $\sigma_b \rightarrow 1$, we have
\begin{equation}\label{eqn:reg}
\lim_{\sigma_b \rightarrow 1}W^*(\delta,G) = n(b(\estar) - c\estar) + cd\delta\estar \sum_{i \in V}x^*_i,
\end{equation}
where $x^*$ denotes the aggregate effort maximizing equilibrium.
 \end{enumerate}
\end{theorem}

The bounds given in \cref{eqn:maxuppbnd}, \cref{eqn:maxlowbnd1} and \cref{eqn:maxlowbnd2} give good approximations to the maximum total welfare so as to help a designer define the structure of interactions to get desirable output. Taking the limit as concavity ($\sigma_b$) tends to 1 in \cref{eqn:maxuppbnd} and \cref{eqn:maxlowbnd2}, we have 
\begin{equation*}
 \lim_{\sigma_b \rightarrow 1} W^*(\delta,G) \geq n(b(\estar) - c\estar) + cd_{min}\delta\estar\alpha(G).
\end{equation*}
 and 
 \begin{equation*}
 \lim_{\sigma_b \rightarrow 1} W^*(\delta,G) \geq n(b(\estar) - c\estar) + cd_{max}\delta\estar\left(\alpha(G)+1 + \frac{1}{\alpha(G)-1}\right).
\end{equation*}

We emphasize once again that the results in this and the following section significantly simplify the problem of the principal of ascertaining maximum aggregate play and welfare over all equilibria. Instead of searching over all equilibria, the principal only needs to calculate certain characteristics of the underlying network, and these suffice to provide good approximations on these aggregate quantities.

\section{Aggregate effort and welfare for tree networks} \label{sec:tree}
An important assumption in the previous section was that $\delta \geq \eta(G).$ We now find explicit formulae for the aggregate effort and welfare for a game where the network does not have cycles (\ie, trees), by only requiring that $\delta \geq \half$.
  
Before we analyze the case of a tree, we first consider simpler cases. We begin with a {\it line network (chain)}. Then we consider {\it starlike} trees which are defined as trees with exactly one node of degree strictly greater than 2. Using these as the building blocks, we give an expression for a tight upper bound on the maximum aggregate effort for general trees based on only the total number of nodes and degrees of each node. On getting the bounds on maximum aggregate effort for trees, we find approximations to welfare using \cref{lem:welbound}.

\subsection{Line networks}

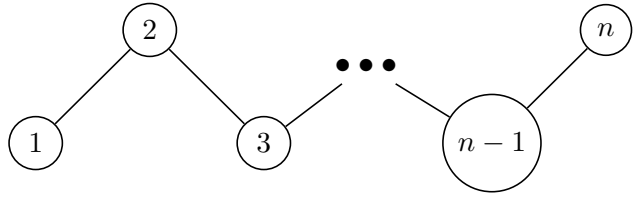
\begin{figure}
 \centering
\begin{tikzpicture}[
  >=stealth,
  semithick,
  every state/.style={draw,circle,font=\sffamily\small, minimum size = 1em}
  ]
 
    \node[state] (1) at (0,0) {$1$};
    \node[state] (2) at (1.5,1.5) {$2$};
    \node[state] (3) at (3,0) {$3$};
    \node[above right =0.75cm of 3] (d) {$\bullet \bullet \bullet$};
    \node[state] (n-1) at (6,0) {$n-1$};
    \node[state] (n) at (7.5,1.5) {$n$};
    
    \draw[-] (1) to  (2);
    \draw[-] (2) to  (3);
    \draw[-] (n-1) to  (n);
    \draw[-] (n-1) to  (d);
    \draw[-] (3) to  (d);
     
\end{tikzpicture}
\caption{A line network with $n$ nodes.}
\label{fig:line_ex}
\end{figure}

In this section, we consider the case of a line network with $n$ nodes and find a tight upper bound on the maximum aggregate play. A sample of such a network is shown in \cref{fig:line_ex}. We also show that when the line network has an odd number of nodes, the upper bound is achieved.

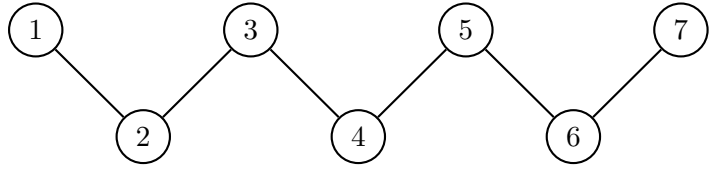
\begin{figure}
\centering
\begin{tikzpicture}[auto, node distance=2cm, every loop/.style={},
                    thick,main node/.style={circle,draw,font=\sffamily\small\bfseries}]

  \node[main node] (1) {$1$};
  \node[main node] (2) [below right of=1] {$2$};
  \node[main node] (3) [above right of=2] {$3$};
  \node[main node] (4) [below right of=3] {$4$};
  \node[main node] (5) [above right of=4] {$5$};
  \node[main node] (6) [below right of=5] {$6$};
  \node[main node] (7) [above right of=6] {$7$};

  \path[every node/.style={font=\sffamily\small}]
    (1) edge node [right] {} (2)
    (2) edge node [right] {} (3)
    (3) edge node [right] {} (4)
    (4) edge node [right] {} (5)
    (5) edge node [right] {} (6)
    (6) edge node [right] {} (7);
\end{tikzpicture} 
\caption{A line network with 7 nodes for Example \ref{ex:line}}
\label{fig:exline}
\end{figure}

\begin{theorem}\label{thm:line}
If $G$ is a line network with $n$ nodes and $\delta \geq \frac{1}{2}$, the maximum aggregate play amongst Nash equilibria is at most $\frac{n+1}{2}\estar$ and is at least $\frac{n}{2}\estar$ , \ie 
$$\frac{n}{2}\estar \leq E^{*}(\delta,G) \leq \frac{n+1}{2}\estar.$$ 
Moreover, the upper bound is achieved for line networks with odd number of nodes.
\end{theorem}
We illustrate all equilibria of a line network of 7 nodes with an example given in \cref{fig:exline}.

\begin{table}
\begin{center}
  \begin{tabular}{ | c || c | c | c | c | c |}
    \hline
    Agent $\downarrow$ & $x^{\RNum{1}}$ & $x^{\RNum{2}}$ & $x^{\RNum{3}}$ & $x^{\RNum{4}}$ & $x^{\RNum{5}}$\\ \hline
    \hline
    1 & $\estar$ & $-\frac{ - \delta^3 + 2\, \delta^2 + \delta - 1}{2\, \delta^4 - 4\, \delta^2 + 1}$ & $\frac{1}{1 + \delta}\estar$ & $\frac{1}{ - \delta^2 + \delta + 1}\estar$ & $\frac{1}{1 + \delta}\estar$\\ \hline
    
    2 & $0$ & $-\frac{ - 2\, \delta^3 + \delta^2 + 2\, \delta - 1}{2\, \delta^4 - 4\, \delta^2 + 1}$ & $\frac{1}{1 + \delta}\estar$ & $\frac{1 - \delta}{ - \delta^2 + \delta + 1}\estar$ & $\frac{1}{1 + \delta}\estar$\\ \hline
    
    3 & $\estar$ & $\frac{\left(\delta - 1\right)\, \left(\delta^2 + \delta - 1\right)}{2\, \delta^4 - 4\, \delta^2 + 1}$ & $0$ & $\frac{1 - \delta}{ - \delta^2 + \delta + 1}\estar$ & $0$\\ \hline
    
    4 & $0$ & $-\frac{2\, \delta - 1}{2\, \delta^4 - 4\, \delta^2 + 1}$ & $\estar$ & $\frac{1}{ - \delta^2 + \delta + 1}\estar$ & $\frac{1}{ - \delta^2 + \delta + 1}\estar$\\ \hline
    
    5 & $\estar$ & $\frac{\left(\delta - 1\right)\, \left(\delta^2 + \delta - 1\right)}{2\, \delta^4 - 4\, \delta^2 + 1}$ & $0$ & $0$ & $\frac{1 - \delta}{ - \delta^2 + \delta + 1}\estar$\\ \hline
    
    6 & $0$ & $-\frac{ - 2\, \delta^3 + \delta^2 + 2\, \delta - 1}{2\, \delta^4 - 4\, \delta^2 + 1}$ & $\frac{1}{1 + \delta}\estar$ & $\frac{1}{1 + \delta}\estar$ & $\frac{1 - \delta}{ - \delta^2 + \delta + 1}\estar$\\ \hline
    
    7 & $\estar$ & $-\frac{ - \delta^3 + 2\, \delta^2 + \delta - 1}{2\, \delta^4 - 4\, \delta^2 + 1}$ & $\frac{1}{1 + \delta}\estar$ & $\frac{1}{1 + \delta}\estar$ & $\frac{1}{ - \delta^2 + \delta + 1}\estar$\\ \hline
    \hline
    Validity $\rightarrow$ & $\delta \in [0.5,1]$ & $\delta \in [\frac{\sqrt{5} - 1}{2},1]  \cup [0,0.5]$ & $\delta \in [\frac{\sqrt{5} - 1}{2},1]$ & $\delta \in [\frac{1}{\sqrt{2}},1]$ & $\delta \in [\frac{1}{\sqrt{2}},1]$ \\ \hline
    \hline
    Total $\rightarrow$ & $4\estar$ & $\frac{8\delta^3 - 6 \delta^2 - 12 \delta + 7}{2\, \delta^4 - 4\, \delta^2 + 1}\estar$ & $\estar + \frac{4}{1+\delta}\estar$ & $\frac{6+4\delta - 4\delta^2}{1 + \delta - \delta^2}\estar$ & $\frac{6+4\delta - 4\delta^2}{1 + \delta - \delta^2}\estar$\\ \hline
  \end{tabular}
\end{center}
    \caption{The validity row shows the range of $\delta$ for which the equilibrium holds. A table showing all equilibria for a line network of 7 nodes.}
    \label{tbl:line}    
\end{table}

\begin{examplec}\label{ex:line}
We consider the line network with $7$ nodes in \cref{fig:exline}. Let $x = (x_1,x_2,\dots,x_7)$ denote a vector of plays. \cref{tbl:line} shows all the equilibria and corresponding aggregate effort of each of them.

From \cref{thm:line}, the aggregate play is upper bounded by $\frac{n+1}{2}\estar = 4\estar$, for $\delta \geq \frac{1}{2}$. {\it Equilibrium I} \ has aggregate play $4\estar$ and hence it is the aggregate play maximizing equilibrium. From Proposition 2 of \cite{bramoulle2014strategic}, the game has a unique Nash equilibrium  for $\delta \leq \frac{1}{|\lambda_{min}(A)|}$ and $\frac{1}{|\lambda_{min}(A)|}  = 0.5412$ in our case. Thus, for $\delta \leq 0.5412$, we have a unique equilibrium. Specifically, for $\delta \leq \frac{1}{2}$, the only equilibrium is {\it Equilibrium II}.
\end{examplec}

\subsection{Stars and Starlike Networks}

\begin{figure}
 \centering
\begin{tikzpicture}[
  >=stealth,
  semithick,
  every state/.style={draw,circle,font=\sffamily\small, minimum size = 1em}
  ]
 
    \node[state] (0) at (0,0) {$0$};
    \node[state] (1) at (2.5,0) {$1$};
    \node[state] (2) at (1.58,1.58) {$2$};
    \node[state] (3) at (0,2.5) {$3$};
    \node[state] (4) at (-1.58,1.58) {$4$};
    \node[state] (5) at (-2.5,0) {$5$};
    \node[state] (6) at (-1.58,-1.58) {$6$};
    \node[right = 0.70cm of 6] (d) {$\bullet \bullet \bullet$};
    \node[state] (n) at (1.58,-1.58) {$n$};
    
    \draw[-] (0) to  (1);
    \draw[-] (0) to  (2);
    \draw[-] (0) to  (3);
    \draw[-] (0) to  (4);
    \draw[-] (0) to  (5);
    \draw[-] (0) to  (6);
    \draw[-] (0) to  (n);
    \draw[-] (0) to  (d);
     
\end{tikzpicture}
\caption{A star with $n$ peripheral nodes.}
\label{fig:star_ex}
\end{figure}
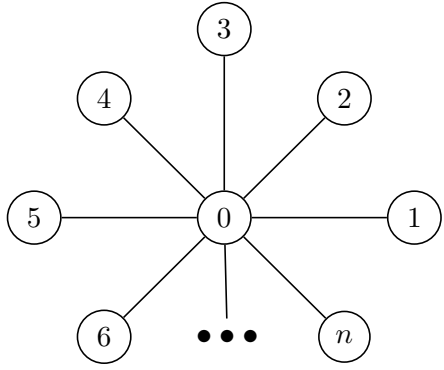

In this section, we consider the case when the interaction structure is a star-like network. First, we show that, for a star with a central node (degree $\geq$ 2) and $n$ peripheral nodes (as shown in \cref{fig:star_ex}), there is a unique equilibrium. 

\begin{theorem}\label{thm:star}
 If $G$ is a star network with $n$ ($\geq 3$) peripheral nodes and $\delta \geq \frac{1}{n}$, the game has only one equilibrium in which all the peripheral nodes play action $\estar$ and the central node free-rides. Thus, the maximum aggregate effort is $n\estar$.
\end{theorem}

Next, we consider a starlike network structure with peripherals coming out of the central node (degree $\geq 3$) themselves being chains instead of just single nodes. We consider $m$ chains $S_1,S_2,\dots,S_m$ with $n_1,n_2,\dots,n_m$ nodes respectively having one leaf node connected to a central node $c$ and the other leaf node free. We find a tight upper bound on the maximum aggregate play for such networks and also show that when all of $n_1,n_2,\dots,n_m$ are odd, \ie all peripheral chains have an odd number of nodes, the upper bound is achieved.

\begin{theorem}\label{thm:starline}
Let $G$ be a star network having $deg_G(c) = m\ (\geq 3)$ chains connected at a central node $c$ and for $\delta \geq \frac{1}{2}$ and let the $r$ out of the $m$ chains have odd number of nodes. Then, the maximum aggregate play amongst Nash equilibria is at most $\frac{|V(G)|+ m - 1}{2}\estar$ and at least  $\frac{|V(G)|+ r - 1}{2}\estar$ \ie 
$$\frac{|V(G)|+ r - 1}{2}\estar \leq E^*(\delta,G) \leq \frac{|V(G)|+ m - 1}{2}\estar.$$
 Moreover, the upper bound is achieved in the case of all chains are of odd length.
\end{theorem}

\begin{figure}
\centering
\begin{tikzpicture}[auto, node distance=2cm, every loop/.style={},
                    thick,main node/.style={circle,draw,font=\sffamily\small\bfseries}]

  \node[main node] (1) {$1$};
  \node[main node] (2) [below right of=1] {$2$};
  \node[main node] (3) [above right of=2] {$3$};
  \node[main node] (4) [below right of=3] {$4$};
  \node[main node] (5) [above right of=4] {$5$};
  \node[main node] (6) [below of=3] {$6$};
  \node[main node] (7) [below right of=6] {$7$};

  \path[every node/.style={font=\sffamily\small}]
    (1) edge node [right] {} (2)
    (2) edge node [right] {} (3)
    (3) edge node [right] {} (4)
    (4) edge node [right] {} (5)
    (3) edge node [right] {} (6)
    (6) edge node [right] {} (7);
\end{tikzpicture} 
\caption{A starlike networks with 7 nodes for Example \ref{ex:starline}}
\label{fig:exstarline}
\end{figure}
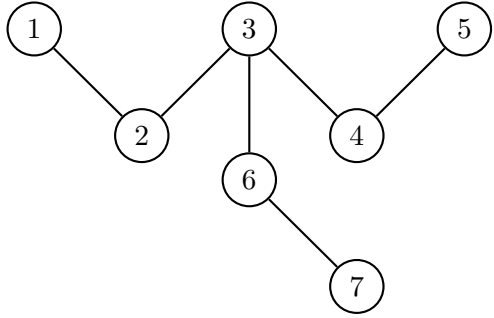

\begin{table}
\begin{center}
  \begin{tabular}{ | c || c | c | c |}
    \hline
    Agent $\downarrow$ & $x^{\RNum{1}}$ & $x^{\RNum{2}}$ & $x^{\RNum{3}}$\\ \hline
    \hline
    1 & $\estar$ & $\frac{\delta + 1}{2\delta + 1}$ & $\frac{1}{1 + \delta}\estar$\\ \hline
    
    2 & $0$ & $\frac{1}{2\delta + 1}$ & $\frac{1}{1 + \delta}\estar$ \\ \hline
    
    3 & $\estar$ & $\frac{1 - \delta}{2\delta + 1}$ & $0$\\ \hline
    
    4 & $0$ & $\frac{1}{2\delta + 1}$ & $\frac{1}{1 + \delta}\estar$ \\ \hline
    
    5 & $\estar$ & $\frac{\delta + 1}{2\delta + 1}$ & $\frac{1}{1 + \delta}\estar$\\ \hline
    
    6 & $0$ & $\frac{1}{2\delta + 1}$ & $\frac{1}{1 + \delta}\estar$ \\ \hline
    
    7 & $\estar$ & $\frac{\delta + 1}{2\delta + 1}$ & $\frac{1}{1 + \delta}\estar$ \\ \hline
    \hline
    Validity $\rightarrow$ & $\delta \in [\frac{1}{2},1]$ & $\delta \in [0,1]$ & $\delta \in [\frac{1}{2},1]$ \\ \hline
    \hline
    Total $\rightarrow$ & $4\estar$ & $\frac{2 \delta + 7}{2\delta + 1}\estar$ & $\frac{6}{1+\delta}\estar$\\ \hline
  \end{tabular}
\end{center}
    \caption{A table showing all equilibria for a starline network of 7 nodes.}
        \label{tbl:starline}
\end{table}

\begin{examplec}\label{ex:starline}
As an example, we consider the network in \cref{fig:exstarline} which is a starlike network with $7$ nodes. Let $x = (x_1,x_2,\dots,x_7)$ denote a vector of plays. \cref{tbl:starline} shows all the equilibria and corresponding aggregate play of each of them.

From \cref{thm:starline}, the aggregate play is upper bounded by $\frac{|V(G)|+m-1}{2} = \frac{7+3-1}{2} = 4.5\estar$, for $\delta \geq \frac{1}{2}$. On comparing the aggregate play of the three equilibria for $\delta \geq \frac{1}{2}$, we get $x^{\RNum{1}}$ is the aggregate play maximizing equilibrium. We also note that the support of $x^{\RNum{1}}$ forms a unique maximum independent set of G, which indicates it being aggregate effort maximizing in $[\eta(G),1)$.  From \cite[Prop. 2]{bramoulle2014strategic}, the game has a unique Nash equilibrium  for $\delta \leq \frac{1}{|\lambda_{min}(A)|}$ and $\frac{1}{|\lambda_{min}(A)|}  = 0.5$ in our case. Thus, for $\delta \leq 0.5$, we have a unique equilibrium. Speicifically, for $\delta \leq \frac{1}{2}$, the only equilibrium is $x^{\RNum{2}}$ and hence it is the aggregate play maximizing equilibrium.
\end{examplec}

\subsection{Trees}

For a general tree $T$, we define the \textit{centers} of the tree as those nodes with degree $\geq 3$. Let the centers of a network $T$ be $\C_T = \{c_1,c_2,\dots, c_t\}$ and let the degree of center $c_i$ in tree $T$ be denoted by $d^T_i$. We define a \textit{branch of a tree} as a path that starts at a center and ends either at a leaf or at another center without passing through another center. We denote the set of branches of a tree by $\B_T$.
 In the next theorem, we give a bound on the aggregate play of any Nash equilibrium of the game on a tree.

\begin{figure}
 \centering
\begin{tikzpicture}[auto, node distance=2cm, every loop/.style={},
                    thick,main node/.style={circle,draw,font=\sffamily\small\bfseries}]
 
   \node[main node] (1) {$1$};
  \node[main node] (2) [below right of  = 1]{$2$};
  \node[main node] (3) [above right of=2] {$3$};
  \node[main node] (4) [below right of=3] {$4$};
  \node[main node] (5) [above right of=4] {$5$};
  \node[main node] (6) [below right of=5] {$6$};
  \node[main node] (7) [above right of=6] {$7$};
  \node[main node] (8) [above right of=5] {$8$};
  \node[main node] (9) [below left of=2] {$9$};
  \node[main node] (10) [below of=3] {$10$};
  \node[main node] (11) [below left of=10] {$11$};
  \node[main node] (12) [below right of=10] {$12$};

  \path[every node/.style={font=\sffamily\small}]
    (1) edge node [right] {} (2)
    (2) edge node [right] {} (3)
    (3) edge node [right] {} (4)
    (4) edge node [right] {} (5)
    (5) edge node [right] {} (6)
    (6) edge node [right] {} (7)
    (5) edge node [right] {} (8)
    (3) edge node [right] {} (10)
    (10) edge node [right] {} (11)
    (9) edge node [right] {} (2)
    (10) edge node [right] {} (12);  
\end{tikzpicture}
\caption{An example tree $T$}
\label{fig:tree_ex}
\end{figure}
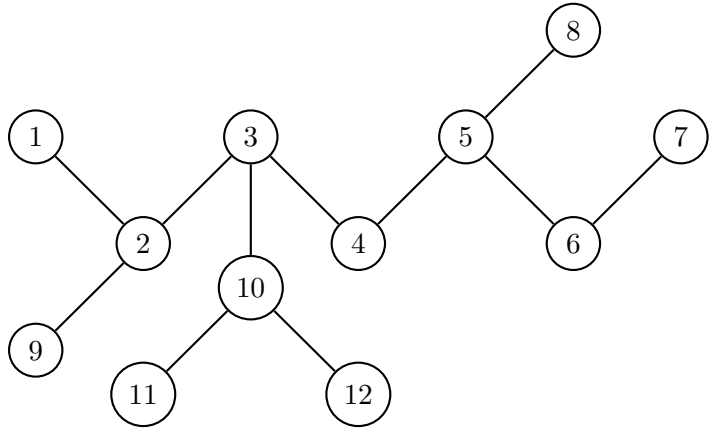

To illustrate the concept of a tree (say $T$), its centers ($\C_T$) and its branches ($\B_T$), we provide an example tree shown in \cref{fig:tree_ex}. The central nodes are $\C_T = \{2,10,3,5\}$ and its branches are $\B_T = \{(2,1),(2,9),(10,11),(10,12),$ $(3,4,5),(5,8),(5,6,7)\}$.

\begin{theorem}\label{thm:tree}
Let $G$ be a tree ($T$) with $\C_T =\{c_1,c_2,\dots, c_t\}$ as the set of centers $(\C_T \neq \emptyset)$ and let $\delta \geq \frac{1}{2}$. Let the total number of branches of $T$ be denoted by $m ( = |\B_T|)$ and let $r$ be the number of odd length branches. Then, the maximum aggregate play amongst Nash equilibria is at most $\frac{|V(G)| + m - |\C_T|}{2}\estar$  and at least $\frac{|V(G)| + r - |\C_T|}{2}\estar$,\ie 
$$\frac{|V(G)| + r - |\C_T|}{2}\estar \leq E^*(\delta,G) \leq \frac{|V(G)| +  m - |\C_T|}{2}\estar.$$
Moreover, the bound is tight in the case when all branches have an odd number of nodes.
\end{theorem}

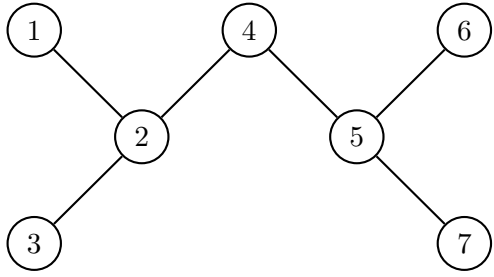
\begin{figure}
\centering
\begin{tikzpicture}[auto, node distance=2cm, every loop/.style={},
                    thick,main node/.style={circle,draw,font=\sffamily\small\bfseries}]

  \node[main node] (1) {$1$};
  \node[main node] (2) [below right of=1] {$2$};
  \node[main node] (3) [below left of=2] {$3$};
  \node[main node] (4) [above right of=2] {$4$};
  \node[main node] (5) [below right of=4] {$5$};
  \node[main node] (6) [above right of=5] {$6$};
  \node[main node] (7) [below right of=5] {$7$};

  \path[every node/.style={font=\sffamily\small}]
    (1) edge node [right] {} (2)
    (2) edge node [right] {} (3)
    (2) edge node [right] {} (4)
    (4) edge node [right] {} (5)
    (5) edge node [right] {} (6)
    (5) edge node [right] {} (7);
\end{tikzpicture} 
\caption{A tree with tight bounds for Example \ref{ex:treetight}}
\label{fig:extreetight}
\end{figure}

\begin{table}
\begin{center}
  \begin{tabular}{ | c || c | c |}
    \hline
    Agent $\downarrow$ & $x^{\RNum{1}}$ & $x^{\RNum{2}}$ \\ \hline
    \hline
    1 & $\estar$ & $\frac{\delta^2 + \delta - 1}{4\, \delta^2 - 1}$ \\ \hline
    
    2 & $0$ & $\frac{3\, \delta - 1}{4\, \delta^2 - 1}$  \\ \hline
    
    3 & $\estar$ & $\frac{\delta^2 + \delta - 1}{4\, \delta^2 - 1}$ \\ \hline
    
    4 & $\estar$ & $-\frac{2\, \delta^2 - 2\, \delta + 1}{4\, \delta^2 - 1}$  \\ \hline
    
    5 & $0$ & $\frac{3\, \delta - 1}{4\, \delta^2 - 1}$ \\ \hline
    
    6 & $\estar$ & $\frac{\delta^2 + \delta - 1}{4\, \delta^2 - 1}$  \\ \hline
    
    7 & $\estar$ & $\frac{\delta^2 + \delta - 1}{4\, \delta^2 - 1}$  \\ \hline
    \hline
    Validity $\rightarrow$ & $\delta \in [\frac{1}{3},1]$ & $\delta \in [0,\frac{1}{3}]$ \\ \hline
    \hline
    Total $\rightarrow$ & $5\estar$ & $\frac{2 \delta^2 + 12\delta - 7}{4\delta^2 - 1}\estar$ \\ \hline
  \end{tabular}
\end{center}
    \caption{A table showing all equilibria for a tree network of 7 nodes.}
    \label{tbl:tree}
\end{table}

\begin{examplec}\label{ex:treetight}
As an example, we consider the network in \cref{fig:extreetight} which is a line network with $7$ nodes. Let $x = (x_1,x_2,\dots,x_7)$ denote a vector of plays. \cref{tbl:tree} shows all the equilibria and corresponding aggregate play of each of them.

Note that for this example, centers are $\C_T = \{2,5\}$ and the branches are $\B_T = \{(2,1),(2,3),$ $(5,6),(5,7),(2,4,5)\}$. From \cref{thm:tree}, the aggregate play is at most $\frac{|V(G)|+ m - |\C_T|}{2}\estar = \frac{7+2 + 2 -1}{2}\estar = 5\estar$, for $\delta \geq \frac{1}{2}$. On comparing the aggregate play of the three equilibria for $\delta \geq \frac{1}{2}$, we get $x^{\RNum{1}}$ is the aggregate play maximizing equilibrium. From Proposition 2 of \cite{bramoulle2014strategic}, the game has a unique Nash equilibrium  for $\delta \leq \frac{1}{|\lambda_{min}(A)|}$ and $\frac{1}{|\lambda_{min}(A)|}  = 0.5$ in our case. Thus, for $\delta \leq 0.5$, we have a unique equilibrium. Speicifically, for $\delta \leq \frac{1}{3}$, the only equilibrium is $x^{\RNum{2}}$ and hence it is aggregate play maximizing. Also, for $\delta \in [\frac{1}{3},\frac{1}{2}]$, the only equilibrium is $x^{\RNum{1}}$ and thus it is aggregate play maximizing.

\end{examplec}

In the next theorem, using the bounds on aggregate play found for trees, we find bounds on the welfare for trees.
\begin{theorem}\label{thm:treewelapprox}
 Consider a strategic interactions game on a tree $T$ with maximum degree $d_{max}$, centers $\C_T =\{c_1,c_2,\dots, c_t\}$ $(\C_T \neq \emptyset)$ and substitutability factor $\delta \geq \frac{1}{2}$. Let the number of branches of $T$ be $m$ and the number of odd branches be $r$. Then, we have the following:
\begin{enumerate}[label = (\alph*)]
\item For $\sigma_b \in (0,1)$, we have
\begin{equation*}
 W^*(\delta,G) \leq n(b(\estar) - c\estar) + cd_{max}\delta\sigma_b\estar\left(\frac{|V(G)| + m - |\C_T|}{2}\right).
\end{equation*}
 \item  For $\sigma_b \leq \frac{1}{1+\delta}$, we have
\begin{equation*}
  W^*(\delta,G) \geq n(b(\estar) - c\estar) + c((\delta + 1)\sigma_b - 1)\estar\left(\frac{|V(G)| + m - |\C_T|}{2}\right).
\end{equation*}
\item  For $\sigma_b \geq \frac{1}{1+\delta}$, we have
\begin{equation*}
  W^*(\delta,G) \geq n(b(\estar) - c\estar) + c((\delta + 1)\sigma_b - 1)\estar\left(\frac{|V(G)| + r - |\C_T|}{2}\right).
\end{equation*}
\end{enumerate}

\end{theorem}

The results of \cref{thm:treewelapprox} give us approximations of the social welfare in the case the underlying network is a tree which depend only on a few characteristics of the underlying network. These characteristics, namely the number of nodes, the number of branches, number of centers are easily computable. We re-emphasize that, thanks to this theorem, the problem of estimating the social welfare has become easier than computing all equilibria and searching over all of them.

\section{Conclusion} \label{sec:concl}

In this paper, we studied games on networks with local contributions. The model is a natural generalization from Bramoull\'e and Kranton's (\cite{bramoulle2007public}) public goods game obtained by the introduction of a substitutability factor. We provide results that aid a system planner in assessing aggregate outcomes and the network underlying the game to achieve desirable outcomes in terms of social welfare.

Specifically, we show that under suitable assumptions on the substitutability factors, the maximum aggregate effort is achieved by an independent clique equilibrium (ICE) and give close approximations to the maximum aggregate play and welfare in terms of graph characteristics. For trees, we provide approximations which are independent of the independence number and depend only on the total number of vertices, degrees of each node of the graph and the structure of the tree. These approximations provide a social planner with a hand on what outcome can be expected if the network underlying the game is chosen in a certain way.

Our work opens up directions for research on how the underlying graph may be edited for ensuring desirable outcomes. Specifically, it provides guidelines on devising changes to the network structure that would result in increases in welfare. It identifies maximum and minimum degrees and independence number of the network as key characteristics that affect the welfare. Thus one may surmise that changes to the network that affect these characteristics are most effective in enhancing the welfare. A formal study of this topic is part of our ongoing work.

\appendix

\begin{section}{Proofs}

\begin{subsection}{Proof of \cref{thm:absmax}}
From \cref{thm:decrplay}, the maximum aggregate effort for any game with $\delta \leq 1$ will be at least as much as that in the case when $\delta = 1$. For $\delta = 1$, we know from \cite{pandit2018refinement} that  the maximum aggregate effort is $\estar \alpha(G)$. Thus, the maximum aggregate play for $\delta \leq 1$ at least $\estar\alpha(G)$.

From \cref{thm:maxics}, we know that the maximum aggregate effort is achieved by an independent clique equilibrium. We find an upper bound on the aggregate effort of any independent clique equilibrium to get a upper bound on the maximum aggregate effort. Let $x$ be an independent clique equilibrium such that $\sigma(x) = \cup_{i = 1}^{\alpha(G)}C_i$, where $C_i$ are independent cliques with $|V(C_i)| = n_i$. Then, the aggregate effort of $x$ is given by
\begin{equation}\label{eqn:aggplayics}
 \sum_{i \in V}x_i = \sum_{i = 1}^{\alpha(G)}\frac{n_i}{1+(n_i-1)\delta}\estar.
\end{equation}

Each term in the RHS of \cref{eqn:aggplayics} is increasing in $n_i$ and decreasing in $\delta$. Thus, to find the maximum value, we take all $n_i = \omega(G)$, which is the size of the largest clique in the network. We know that $\delta \geq \frac{\alpha(G)(\omega(G)-1) - \omega(G)}{\alpha(G)(\omega(G)-1)}$, so we take $\delta = \frac{\alpha(G)(\omega(G)-1) - \omega(G)}{\alpha(G)(\omega(G)-1)}$ to find the bound. Substituting in \cref{eqn:aggplayics}, we get 
\begin{align*}
  \sum_{i \in V}x_i &=  \sum_{i = 1}^{\alpha(G)}\frac{n_i}{1+(n_i-1)\delta}\estar\\
  & \leq \frac{\alpha(G)\omega(G)}{1+(\omega(G)-1)\frac{\alpha(G)(\omega(G)-1) - \omega(G)}{\alpha(G)(\omega(G)-1)}}\estar\\
  & = \frac{\alpha(G)\omega(G)}{\frac{\alpha(G) + \alpha(G)\omega(G) - \alpha(G) - 1}{\alpha(G)}}\estar\\
  & = \frac{\alpha^2(G)}{\alpha(G)-1}\estar.
\end{align*}
Thus, we have $\sum_{i \in V}x_i \leq \left(\alpha(G)+1 + \frac{1}{\alpha(G)-1}\right)\estar$. Since this is true for all independent clique equilibria with $\alpha(G)$ cliques, we have $E^*(\delta,G) \leq \left(\alpha(G)+1 + \frac{1}{\alpha(G)-1}\right)\estar$. \hfill $\blacksquare$
\end{subsection}~

\subsection{Proof of \cref{lem:welbound}}
We first prove\cref{eqn:welbound1}. From the first inequality in\cref{eqn:linapprox}, we have
\begin{align}
  W(x;\delta,G) \nonumber
  &= \sum_{i \in V}(b(\C_i(x)) -c x_i)\\\nonumber
  &\geq \sum_{i \in V}(b(\estar) - c\estar + c(\sigma_b\C_i(x) - x_i))\\ \nonumber
  &= n(b(\estar) - c\estar) + c\left(\delta\sigma_b\sum_{i \in V}\sum_{j \in V}a_{ij}x_j - (1 - \sigma_b)\sum_{i \in V}x_i\right)\\ \label{eqn:welbound1_last}
  &\geq n(b(\estar) - c\estar) + c((d_{min}\delta + 1)\sigma_b - 1)\sum_{i \in V} x_i. 
\end{align}
Here, \cref{eqn:welbound1_last} follows from the fact that $\sum_{i \in V}\sum_{j \in V}a_{ij}x_j = \sum_{i \in V}d_ix_i \geq d_{min}\sum_{i \in V}x_i$, where $d_i$ denotes the degree of node $i$. Thus \cref{eqn:welbound1} is proved.

Next, we prove \cref{eqn:welbound2} using the second inequality in \cref{eqn:linapprox}.
\begin{align}
  W(x;\delta,G) \nonumber
  &= \sum_{i \in V}(b(\C_i(x)) -c x_i)\\\nonumber
  &\leq \sum_{i \in V}(b(\estar) - c\estar + c(\C_i(x) - x_i))\\ \nonumber
  &= n(b(\estar) - c\estar) + c\delta\sum_{i \in V}\sum_{j \in V}a_{ij}x_j\\ \label{eqn:welbound2_last}
  &\leq n(b(\estar) - c\estar) + c(d_{max}\delta\sigma_b)\sum_{i \in V} x_i. 
\end{align}
Here, \cref{eqn:welbound2_last} follows from $\sum_{i \in V}\sum_{j \in V}a_{ij}x_j = \sum_{i \in V}d_ix_i \leq d_{max}\sum_{i \in V}x_i$, where $d_i$ denotes the degree of node $i$. Thus \cref{eqn:welbound2} is proved. \hfill $\blacksquare$

\subsection{Proof of \cref{thm:welapprox}}
 Let the aggregate play maximizing ICE be given by $x^*$. 
 \begin{enumerate}[label = (\alph*)]
  \item We maximize the RHS of \cref{eqn:welbound2} over $x$ to get 
\begin{align}\nonumber
 W(x;\delta,G) &\leq  n(b(\estar) - c\estar) + cd_{max}\delta\sigma_b\sum_{i \in V} x^*_i \\ \label{eqn:maxuppbndstep}
 &\leq n(b(\estar) - c\estar) + cd_{max}\delta\sigma_b\estar\left(\alpha(G)+1 + \frac{1}{\alpha(G)-1}\right)
\end{align}
\eqref{eqn:maxuppbndstep} follows from \cref{thm:absmax}. Since \cref{eqn:maxuppbndstep} holds for every $x$ such that $\frac{x}{\estar} \in \SOL_\delta(G)$, we have 
\begin{equation*}
 W^*(\delta,G) \leq n(b(\estar) - c\estar) + cd_{max}\delta\sigma_b\estar\left(\alpha(G)+1 + \frac{1}{\alpha(G)-1}\right).
\end{equation*}
 \item Since \cref{eqn:welbound1} holds for all $x$, we consider this equation for $x^*$, \ie 
\begin{equation*}
  W^*(\delta,G) \geq W(x^*;\delta,G) \geq n(b(\estar) - c\estar) + c((d_{min}\delta + 1)\sigma_b - 1)\sum_{i \in V} x^*_i
\end{equation*}
For $\sigma_b \leq \frac{1}{1+d_{min}\delta}$, to minimize the lower bound, we need to consider the maximum value of $\sum_{i \in V}x^*_i$. Thus, we have for $\sigma_b \leq \frac{1}{1+d_{min}\delta}$

\begin{equation*}
  W^*(\delta,G) \geq n(b(\estar) - c\estar) + c((d_{min}\delta + 1)\sigma_b - 1)\estar\left(\alpha(G)+1 + \frac{1}{\alpha(G)-1}\right).
\end{equation*}
\item For $\sigma_b \geq \frac{1}{1+d_{min}\delta}$, to minimize the lower bound, we need to consider the minimum value of $\sum_{i \in V}x^*_i$.
Thus, we have for $\sigma_b \geq \frac{1}{1+d_{min}\delta}$

\begin{equation*}
  W^*(\delta,G) \geq n(b(\estar) - c\estar) + c((d_{min}\delta + 1)\sigma_b - 1)\estar\alpha(G).
\end{equation*}
\item In the case when unique maximum independent sets exist, \cref{thm:uniqmax} dictates that the maximum aggregate play is $\estar \alpha(G)$. Using this value of $\sum_{i = 1 \in V}x^*_i$ in the proof of \cref{thm:welapprox}, we get the bounds as
\begin{equation*}
 W^*(\delta,G) \leq n(b(\estar) - c\estar) + cd_{max}\delta\sigma_b\estar \alpha(G)
\end{equation*}
and 
\begin{equation*}
  W^*(\delta,G) \geq n(b(\estar) - c\estar) + c((d_{min}\delta + 1)\sigma_b - 1)\estar \alpha(G).
\end{equation*}

\item Notice that in the claim, $\sigma_b$ is varied keeping $b(e^*)$ and $b'(e^*)$ fixed. Consequently, the set of equilibria of the game does not change with $\sigma_b$ (see~\cite{pandit2018refinement} for examples of such a sequence of benefit functions), and hence $x^*$ does not depend on $\sigma_b$. 
Taking the limit as $\sigma_b \rightarrow 1$ in \cref{eqn:welbound1} we get, 
 \[
  \lim_{\sigma_b \rightarrow 1} W(x;\delta,G)
  \geq n(b(\estar) - c\estar) + cd\delta\sum_{i \in V} x_i.
  \]
  Thus, 
  \begin{equation*}
  \lim_{\sigma_b \rightarrow 1}W^*(\delta,G) \geq \lim_{\sigma_b \rightarrow 1}W(x^*;\delta,G) \geq n(b(\estar) - c\estar) + cd\delta\sum_{i \in V} x^*_i.
\end{equation*}
Next, taking the maximum over all $x$ which are equilibria in \cref{eqn:welbound2} we get,
\[
 W^*(\delta,G) \leq  n(b(\estar) - c\estar) + cd\sigma_b\delta\sum_{i \in V} x^*_i,
\]
where $x^*$ is the aggregate effort maximizing equilibrium. Similarly for the upper bound we have, 
\[
 \lim_{\sigma_b \rightarrow 1}W^*(\delta,G) \leq  n(b(\estar) - c\estar) + cd\delta\sum_{i \in V} x^*_i,
\]
giving
 \[
  \lim_{\sigma_b \rightarrow 1}W^*(\delta,G) = n(b(\estar) - c\estar) + cd\delta\sum_{i \in V}x^*_i, 
 \]
 as required. 
 \hfill $ \blacksquare$
 \end{enumerate}
 
\begin{subsection}{Proof of \cref{thm:line}}
First, we prove the lower bound. From \cref{thm:lower}, we know that $\sum_{i \in V}x_i \geq \alpha \estar$, where $\alpha$ is the independence number of the path. Now, we note that every tree is a bipartite graph and for a bipartite graph with $n$ nodes, we have $\alpha \geq \frac{n}{2}$. Thus, we have  $\sum_{i \in V}x_i \geq \alpha \estar \geq \frac{n}{2}\estar$.

 We will prove the upper bound by induction. For the base case, note that for a line network with just $1$ node, the only equilibrium is itself contributing $\estar$, making the aggregate play $\estar = \frac{1+1}{2}\estar = \frac{n+1}{2}\estar$.

Let the aggregate play be upper bounded by $\frac{k+1}{2}\estar$ for all line networks with number of nodes equal to $k<n$. Let $P_{n}$ denote the line network with $n$ nodes. Let $x$ denote an equilibrium of the game on $P_n$ and $\delta \geq \frac{1}{2}$.

\noindent {\it Case 1:} $\sigma(x) = V.$\\
If an equilibrium has full support the discounted sum of efforts of the closed neighbourhood of each node is $\estar$, \ie $\C_i(x) = \estar$. Then,

\begin{align*}
\sum_{i = 1}^{n} \C_i(x) &= \sum_{i=1}^{n}x_i + \delta\sum_{i=1}^{n}x_i + \delta\sum_{i=1}^{n}x_i -  \underbrace{\delta(x_1 + x_{n})}_{\text{$1$ and $n$ are leaves}}\\
n\estar + 2\estar \delta&\geq (2\delta+1)\sum_{i=1}^{n}x_i \\
\frac{n + 1}{2}\estar & \geq \sum_{i=1}^{n}x_i, \ 
\end{align*}
where the last inequality follows since $\delta \geq \frac{1}{2}.$ 
Thus, the aggregate play of an equilibrium with full support is less than or equal $\frac{n+1}{2}\estar$. 

\noindent {\it Case 2:} $\sigma(x) \subset V$, a strict subset.\\
If the $(r+1)^{th}$ node is assumed to contribute zero, we get two disconnected subnetworks $P_r$ and $P_{n - 1 - r}$. Using \cref{lem:lcp_genprop}, $x_{P_r}$ is an equilibrium of the game with the network as $P_r$ and $x_{P_{n-1-r}}$ is an equilibrium of the game with the network $ P_{n-1-r}$. By induction hypothesis, aggregate play of any equilibrium in $P_r$ is less than or equal to $\frac{r+1}{2}\estar$ and that of any equilibrium in $P_{n - 1 - r}$ is less than or equal to $\frac{n - r}{2}\estar$. Thus the total aggregate play is less than or equal to $\frac{r+1}{2}\estar + \frac{n-r}{2}\estar = \frac{n+1}{2}\estar$. By the principle of mathematical induction, the first part of the theorem is proved.

For the tightness of the bound, consider a line network with $2k+1$ nodes and $x_{2i+1} = \estar \ \forall i \in \{0,1,\dots, k\}$ is an equilibrium with aggregate effort $\frac{2k+1+1}{2}\estar = (k+1)\estar$. Hence, we see that the bound is achieved for line networks with odd number of nodes. \hfill $\blacksquare$
\end{subsection}~

\begin{subsection}{Proof of \cref{thm:star}}
Consider the star network $G$ with $n$ peripheral nodes and let $x$ be an equilibrium of the game with $\delta \geq \frac{1}{n}$. We again consider two cases.

\noindent {\it Case 1:} $\sigma(x) = V.$\\
Then, on solving the $(I + \delta A)x = \bfe\estar$ , we get
\[
x_0 = \frac{n\delta - 1}{n\delta^2-1}\estar \ \text{and} \ x_1 = x_2 = \dots = x_n = \frac{\delta - 1}{n\delta^2-1}\estar
\]

Since $x_0$ and $x_i , i \neq 0$ have opposite signs for all values of $\delta$ with $\delta \geq \frac{1}{n}$, such an equilibrium cannot exist.\\

\noindent {\it Case 2:} $\sigma(x) \subset V$, a proper subset\\
Note that, any vector $x$ with $x_i = 0$ where $i$ is a peripheral node cannot be a Nash equilibrium because the equilibrium conditions can't be satisfied for the peripheral node if it is contributing $0$. Let $x_0 = 0$, where $0$ is the central node. Then the candidate equilibrium $x$ is $x_0 = 0$ and $x_i = \estar \ \forall i \in \{1,2,\dots,n\}$. For this $x$, $\C_0(x) = n\delta \estar \geq \estar$ for $\delta \geq \frac{1}{n}$.  Hence, $x$ is an equilibrium.

Hence, the only equilibrium in a star network is one in which all the peripheral nodes play $\estar$ and the central node free-rides.\hfill $\blacksquare$ 
\end{subsection}

\begin{subsection}{Proof of \cref{thm:starline}}
First, we prove the lower bound. From \cref{thm:lower}, we know that $\sum_{i \in V}x_i \geq \alpha \estar$, where $\alpha$ is the independence number of the tree. Now, we note that for every chain, there exists an independent set with $\frac{n+1}{2}$ nodes when $n$ is odd and there exists an independent set with $\frac{n}{2}$ nodes when $n$ is even, where $n$ is the number of nodes in the chain. We denote the set of chains emanating from $c$ as $\mathcal{P}^c = \{P_1,P_2,\dots,P_m\}$. Then, combining independent sets of every chain in $\mathcal{P}^c$ gives an independent set of the star network $G$. This is because for any pair of nodes $i \in P_i$ and $j \in P_j$ ($i \neq j$), we have $N_G(i) \cap \{j\} = \emptyset$. Let $|P_i| = n_i$  Thus, we have $\alpha \geq \frac{\sum_{n_i  {\rm is}\  {\rm odd}}(n_i + 1) + \sum_{n_j \  {\rm is} \  {\rm even}}(n_j)}{2} = \frac{|V(G)| + r -1}{2}$, since $|V(G)| = \sum_{i = 1}^{m}n_i + 1$. Hence, $\frac{|V(G)|+ r - 1}{2}\estar \leq \sum_{i=1}^{n}x_i$.

We will prove the upper bound by induction on the number of nodes. After showing the result for a base case, we assume that the result holds for all stars smaller than $T$ and show that it holds for $T$. To do this, we divide our proof into different cases depending on what the support of an aggregate effort maximizing equilibrium is. The first case is when it is of full support. Here, we show the bound using some algebraic manipulation. Next, we assume that the central node contributes 0. The result then follows by finding upper bounds on paths using \cref{thm:line} and combining them. The final two cases are when a neighbour of the central node contributes nothing or any other node along the paths contribute 0. Here, we use the induction hypothesis and \cref{thm:line} to show the results.

First, for the base case, consider a star network with 3 peripheral nodes ($m = 3$). The only equilibrium is when each peripheral node contributes $\estar$ and the central node contributes $0$ for $\delta \geq \frac{1}{2}$, and its aggregate play is $ 3\estar = \frac{4+3-1}{2}\estar$. Hence, our bound holds for the base case.

Consider the star $G$ having line networks $S_1,S_2,\dots,S_m$ with $n_1,n_2,\dots,n_m$ nodes respectively. Let $x$ be an equilibrium with the network $G$.\\

\noindent {\it Induction Hypothesis:} Let  $\delta \geq \frac{1}{2}$. Then, for all stars $G'$ with $m'\ ( \geq 3)$ chains connected at a central node $c'$  with $|V(G')| < |V(G)|$, the aggregate play of a Nash equilibrium is at most $\frac{|V(G')|+ m' - 1}{2}$.\\

\noindent {\it Case 1:} $\sigma(x) = V.$\\
Let $x_c$ denote the action of the central node and $x^i_l$ denote the action of the leaf of the line network $S_i$. If an equilibrium has full support the discounted sum of plays of the closed neighbourhood of each node is $\estar$, \ie $\C_i($x$) = \estar$. Then,
\begin{align*}
\sum_{i \in V} \C_i(x) &= \sum_{i\in V}x_i + 2\delta(\sum_{i \in V}x_i) + \delta(x_c) - \delta(\sum_{i = 1}^{m}x^{i}_{l})\\
\frac{|V(G)|\estar + m\delta \estar}{(2\delta + 1)} &\geq \sum_{i\in V}x_i \  \text{ (taking worst case values of $x_c$ and $x^{i}_{l}$)} \\
\frac{|V(G)| + \frac{m}{2}}{2}\estar & \geq \sum_{i \in V}x_i \ \ \ \text{ (since the LHS above is maximized at $\delta$ = 0.5)} \\ 
\frac{|V(G)| + m - 1}{2}\estar & \geq \sum_{i \in V}x_i, 
\end{align*}
where the last inequality is because $\frac{m}{2} \leq m-1.$
Thus, the aggregate play of an equilibrium with full support is less than or equal to $\frac{|V(G)| + m - 1}{2} \estar$. \\

\noindent {\it Case 2:} $\sigma(x) \subset V$, a proper subset.\\
We divide this case into three separate cases. In each case, the on restricting the network to all nodes except one, which contributes to 0, we get a few smaller disconnected subnetworks of the original network. Using \cref{lem:lcp_genprop}, $x$ restricted to these networks is an equilibrium of the game on these networks. Thus, finding upper bounds on the aggregate effort of equilibria on these networks and adding them up gives an upper bound on the aggregate effort of the original network.\\

\noindent {\it Case 2a:} $x_c = 0$.\\
On restricting the network to the rest of the nodes, we get $m$ disjoint line networks of lengths $n_1,n_2,\dots,n_m$. Applying \cref{thm:line} to each line network and adding the upper bounds, we get the upper bound as $\frac{\sum_{i = 1}^{m}n_i + m}{2}\estar = \frac{|V(G)| + m - 1}{2}\estar$. \\

\noindent {\it Case 2b:} $x_k = 0$ where $k$ is the $(r+1)^{th}$ node from $c$, where $n_m-1 \geq r \geq 1$.\\
On restricting the network to the rest of the nodes, we get a star with $m$ line networks and $|V(G)| - n_m + r$ nodes and a line network of length $n_m-1-r$. By the induction hypothesis, since the new star has less number of nodes than $G$, and by \cref{thm:line}, we get the upper bound as $\frac{|V(G)| - n_m + r + m - 1}{2}\estar + \frac{n_m-1-r+1}{2}\estar =\frac{|V(G)| + m - 1}{2}\estar$.\\

\noindent {\it Case 2c:} The neighbour of the central node in $S_m$ contributes 0.\\
On restricting the network to the rest of the nodes, we get a star with $m-1$ peripheral line networks with $|V(G)| - n_m$ nodes and a line network with $n_m-1$ nodes. For $m \geq 4$, by the induction hypothesis, since the new star-like network has fewer nodes than $G$, and by \cref{thm:line}, we get the upper bound as $\frac{|V(G)|+(m-1) -1}{2}\estar + \frac{n_m-1+1}{2}\estar =\frac{|V(G)| + m - 2}{2}\estar \leq \frac{|V(G)| + m - 1}{2}\estar$. For $m = 3$, we get two disjoint line networks with $|V(G)| - n_3$ and $n_3 - 1$ nodes respectively, for which using \cref{thm:line} gives the upper bound as $\frac{|V(G)| - n_3+1}{2}\estar + \frac{n_3-1+1}{2}\estar = \frac{|V(G)| + 1}{2}\estar < \frac{|V(G)| - 1 + m}{2}\estar$.

Hence, the aggregate play is at most $\frac{|V(G)| + m - 1}{2}\estar$ for all equilibria. For the achievability of the bound, consider $x$ defined as follows with line networks of length $2k_1+1,2k_2+1,\dots,2k_m+1$.
\begin{eqnarray*}
 x_c = &0,\\
 x_i^j =&
\begin{cases}
\estar , & {\rm  if } \ j \ {\rm is} \ {\rm odd}, \\
0 , & {\rm  otherwise },
\end{cases}    
\end{eqnarray*}
where $x^j_i$ is the contribution of the $j^{th}$ node of the line network $S_i$ when we start counting from the leaf. It can be seen that $||x||_1 = \sum_{i = 1}^{m} k_i + m = \frac{\sum_{i = 1}^{m}n_i + m}{2}\estar = \frac{|V(G)| - 1 + m}{2}\estar$. $\blacksquare$
\end{subsection}

\begin{subsection}{Proof of \cref{thm:tree}}
First, we prove the lower bound. From \cref{thm:lower}, we know that $E^*(\delta,G) \geq \alpha \estar$, where $\alpha$ is the independence number of the tree. Now, we note that for every chain, there exists an independent set with $\frac{n+1}{2}$ nodes when $n$ is odd and there exists an independent set with $\frac{n}{2}$ nodes when $n$ is even, where $n$ is the number of nodes in the chain. We denote the set of branches as $\B_T$.The union of independent sets of every branch in $\B_T$ gives an independent set of the tree $G$.  This is because for any pair of nodes $i \in B_i$ and $j \in B_j$ ($B_i$ and $B_j$ are distinct chains from $\B_T$), we have $N_G(i) \cap \{j\} = \emptyset$. Thus, we have 
$\alpha \geq \frac{\sum_{B \in \B_T:|B|  {\rm is} \  {\rm odd}}(|B| + 1) + \sum_{B \in \B_T:|B| \  {\rm is} \  {\rm even}}(|B|)}{2} = \frac{|V(G)| + r -|\C_T|}{2}$, since $|V(G)| = \sum_{B \in \B_T}|B| + |\C_T|$. Hence, $\frac{|V(G)|+ r - |\C_T|}{2}\estar \leq E^*(\delta,G)$.

We will prove the upper bound by induction on the number of nodes. We will prove the upper bound by induction on the number of nodes. After showing the result for a base case, we assume that the result holds for all stars smaller than $T$ and show that it holds for $T$. To do this, we divide our proof into different cases depending on what the support of an aggregate effort maximizing equilibrium is. The case when the equilibrium has full support is shown similar to the proof of \cref{thm:starline}. Next, we assume that one of the central nodes contributes 0. The result then follows by finding upper bounds on paths using \cref{thm:line}, using mathematical induction on the disjoint trees formed by restricting the graph to all nodes except the center contributing 0 and combining them. The next two cases are when the neighbour of a central node in a branch terminating in a leaf contributes 0 and when any other node in a branch terminating in a leaf contributes 0. We prove the bound in this case using \cref{thm:line} on a path and the induction hypothesis on a smaller graph. In the last two cases, we consider the case when the neighbour of a central node in a branch terminating in another center contributes 0 and when any other node in a branch terminating in another center contributes 0. The proof in this case involves the use of induction hypothesis on the two smaller trees formed by restricting the graph to all nodes except the one contributing 0.

First, for the base case, consider a star network with 3 peripheral nodes ($m_1 = 3$), \ie $\C_T = {c_1}$, $\B_T = \{\{1\},\{2\},\{3\}\}$. Using \cref{thm:starline}, its aggregate play $\leq 3 = \frac{4+3-1}{2}$. Hence, our bound holds for the base case.

Let $G$ be a tree ($T$) with $\C_T =\{c_1,c_2,\dots, c_t\}$ as the set of centers $(\C_T \neq \emptyset)$. Let $P_{ij}$ denote the branch connecting $c_i$ and $c_j$. Let the degree of center $c_i$ be denoted by $d_i$. Let center $c_i$ have branches $S^i_1,S^i_2,\dots ,S^i_{m_i}$ of length $n^i_1,n^i_2,\dots ,n^i_{m_i}$ connected to it which terminate at leaves.\\

\noindent {\it Induction Hypothesis}: For all trees $T'$ with centers $\C_{T'} = \{c'_1,c'_2,\dots, c'_{t'}\} (\C_T \neq \emptyset)$, $\B_{T'}$ as the set of branches ($|\B_{T'}| = m$) and for $\delta \geq \frac{1}{2}$ with $|V(G')| < |V(G)|$, the aggregate play of a Nash equilibrium is at most $\frac{|V(T')| + m - |\C_T|}{2}$.\\

\noindent {\it Case 1:} $\sigma(x) = V.$\\
Let $x_{c_j}$ denote the action of centre $c_j$ and $x^{j}_{i,l}$ is the action of the leaf of path $S^j_i$. If an equilibrium has full support the discounted sum of plays of the closed neighbourhood of each node is one, \ie $(\C_i($x$)) = 1$. Then,

\begin{align*}
\sum_{i \in V} C_i(x) &= \sum_{i\in V}x_i + 2\delta(\sum_{i \in V}x_i) + \delta(\sum_{c_j \in \C_T}(d_j - 2)x_{c_j}) - \delta(\sum_{c_j \in \C_T}\sum_{i = 1}^{m_j}x^{j}_{i,l}) \\
|V(T)| + \delta(\sum_{c_j \in \C_T}\sum_{i = 1}^{m_j}x^{j}_{l,i}) &= (2\delta + 1)\sum_{i\in V}x_i + \delta(\sum_{c_j \in \C_T}(d_j - 2)x_{c_j}) \\
\frac{|V(T)| + \delta(\sum_{c_j \in \C_T}m_j)}{(2\delta + 1)} &\geq \sum_{i\in V}x_i \text{(putting worst case values of $x_{c_j}$ and $x^j_{l,i}$)} \\
\frac{|V(T)| + \frac{\sum_{c_j \in \C_T}m_j}{2}}{2} & \geq \sum_{i \in V}x_i \text{ (since the LHS above is maximized at $\delta = 0.5$)} \\ 
\frac{|V(T)| + \sum_{c_j \in \C_T}m_j - 1}{2} & \geq \sum_{i \in V}x_i,
\end{align*}
since $\frac{\sum_{c_j \in \C_T}m_j}{2} \leq \sum_{c_j \in \C_T}m_j - 1.$
Now, there are $|\C_T| - 1$ branches which have both endpoints as centers and $m = |\B_T| =  \sum_{c_j \in \C_T}m_j + |\C_T| - 1$. Thus, the aggregate play of an equilibrium with full support is at most $\frac{|V(G)| + m - |\C_T|}{2}$. \\

\noindent {\it Case 2:} $\sigma(x) \subset V$, a proper subset\\
We divide this case into five separate cases. In each case, the on restricting the network to all nodes except one, which contributes to 0, we get a few smaller disconnected subnetworks of the original network. Using \cref{lem:lcp_genprop}, $x$ restricted to these networks is an equilibrium of the game on these networks. Thus, finding upper bounds on the aggregate effort of equilibria on these networks and adding them up gives an upper bound on the aggregate effort of the original network. \\

\noindent {\it Case 2a:} $x_{c_j} = 0$ for some $c_j \in \C_T$.\\

On restricting the network to the rest of the nodes, we get $m_j$ line networks of lengths $n^j_1,n^j_2,\dots,n^j_{m_j}$ and $d_j - m_j = t_j$ disjoint trees $T^1,\dots, T^{t_j}$ with centers $\C_{T^1},\dots, \C_{T^{t_j}}$ such that $ \bigcup_{i = 1}^{t_j} \C_{T^i} \subset \C_T \backslash {c_j}$ and branches $\B_{T^1},\dots, \B_{T^{t_j}}$. By the induction hypothesis, \cref{thm:line} and adding all the bounds we get 
$$\sum_{i \in V}x_i \leq \sum_{i = 1}^{t_j} \frac{|V(T^i)| + |\B_{T^i}| - |\C_{T^i}|}{2} + \frac{\sum_{i = 1}^{m_j}n^j_i + {m_j}}{2}.$$
Note that $\sum_{i = 1}^{t_j}|V(T^{i})| + \sum_{i = 1}^{m_j}n^j_i = |V(T)| - 1$. Also, if $|\C_T| =\sum_{i = 1}^{t_j} |\C_{T^i}| + 1$, no new branches are added and we have $|\B_T| = \sum_{i = 1}^{t_j} |\B_{T^i}| + m_j$, giving $\sum_{i = 1}^{t_j}(|\B_{T^i}| - |\C_{T^i}|) - 1 + m_j = |\B_T| - |\C_T|$. If $|\C_T|  - \sum_{i = 1}^{t_j} |\C_{T^i}| - 1 = k(> 0)$, then $|\B_T| - \sum_{i = 1}^{t_j} |\B_{T^i}| - m_j = k$. This is because whenever a node $c_i\ (\neq c_j)$ which was a center in $T$ but is not a center in $T_i$ formed by restricting $T$ to vertex set $V(T)\backslash \{c_j\}$, the number of branches in $T_i$ also decreases by 1. This is graphically shown in \cref{fig:case_center0}. Thus, we have $\sum_{i = 1}^{t_j}(|\B_{T^i}| - |\C_{T^i}|) - 1 + m_j = |\B_T| - |\C_T|$ giving 
$$\sum_{i \in V}x_i \leq \frac{|V(T)| + m - |\C_T|}{2}.$$
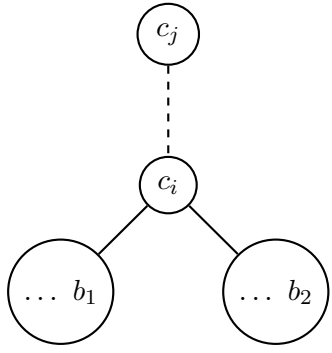
\begin{figure}
 \centering
\begin{tikzpicture}[auto, node distance=2cm, every loop/.style={},
                    thick,main node/.style={circle,draw,font=\sffamily\small\bfseries}, text node/.style={circle,draw,font=\sffamily\small}]
 
   \node[main node] (3) {$c_i$};
  \node[main node] (2) [above of  = 3]{$c_j$};
  \node[text node] (1) [below left of=3] {  $\dots$ $b_1$ };
  \node[text node] (4) [below right of=3] {$\dots$ $b_2$};
  
  \path[every node/.style={font=\sffamily\small}]
    (1) edge node [right] {} (3)
    (2) edge[dashed] node [right] {} (3)
    (3) edge node [right] {} (4);  
\end{tikzpicture}
\caption{$c_j$ is the node with $x_{c_j} = 0$. On restricting $T$ to $V(T) \backslash \{c_j\}$, edge $c_i,c_j$ ceases to exist and $c_i$ is no longer a center in the new disjoint tree. $b_1$ and $b_2$ denote the two branches emanating from $c_i$ in $T$. But, in the new tree, the number of branches also decreases by 1 since $b_1$ and $b_2$ now merge into a single branch including $c_i$.}
\label{fig:case_center0}
\end{figure}

\noindent {\it Case 2b:} The $(r+1)^{th}$ node from $c_j$ in path $S^j_{i}$ contributes $0$, where $n_{m_j}-1 \geq r \geq 1$, \ie$x^j_{i,r+1} = 0$ \\
On restricting the network to the rest of the nodes, we get a path of length $n_{m_j}-1-r$ and a tree $T'$ with centers $\C_{T'} = \C_T$ and $|V(T')| = |V(T)| - (n_{m_j}-r)$ and the number of branches remain the same. By the induction hypothesis and \cref{thm:line}, we get $\sum_{i \in V}x_i \leq \frac{|V(T')| + m - |\C_T|}{2} + \frac{n_{m_j}-1-r + 1}{2} \leq \frac{|V(T)| + m - |\C_T|}{2}$.\\

\noindent {\it Case 2c:} The neighbour of a central node $c_j$ in $S^j_{i}$ contributes 0, \ie $x^j_{i,1} = 0$.\\
On restricting the network to the rest of the nodes, we get a path of length $n_{m_1}-1$ and a tree $T'$ with centers $\C_{T'} = \C_T$, $|V(T')| = |V(T)| - (n_{m_j})$ and $|\B_{T'}| = |\B_T| - 1$. By the induction hypothesis and \cref{thm:line}, we get the upper bound as $\sum_{i \in V}x_i \leq \frac{|V(T')| + |\B_{T'}| - |\C_{T'}|}{2} + \frac{n_{m_j}-1 + 1}{2} = \frac{|V(G)| + |\B_T| - |\C_T| - 1}{2}< \frac{|V(G)| + |\B_T| - |\C_T|}{2}$.\\

\noindent {\it Case 2d:} The $(r+1)^{th}$ node from $c_j$ in $P^{ij}_{join}$ contributes 0, where $|P^{ij}_{join}|-3 > r \geq 1$.\\
On restricting the network to the rest of the nodes, we get two trees $T^i$ and $T^j$ with centers $\C_{T^i},\C_{T^j}$ satisfying $\C_{T^i} \cup \C_{T^j} = \C_T$ and branches $\B_{T^i}, \B_{T^j}$ satisfying $|\B_{T^i}| + |\B_{T^j}| = |\B_T| + 1$. Also, $|V(T^i)| + |V(T^j)| = |V(T)| - 1$. By Induction Hypothesis, we get 
$$\sum_{i \in V}x_i \leq \frac{|V(T^i)| + |\B_{T^i}| - |\C_{T^i}|}{2} +\frac{|V(T^j)| + |\B_{T^j}| - |\C_{T^j}|}{2} = \frac{|V(G)| +  |\B_{T}| - |\C_{T}|}{2}$$.

\noindent {\it Case 2e:} The neighbouring node of $c_j$ in $P^{ij}_{join}$ contributes 0.\\
On restricting the network to the rest of the nodes, we get two trees $T^i$ and $T^j$ with centers $\C_{T^i},\C_{T^j}$ and branches $\B_{T^i}, \B_{T^j}$ respectively. Note that $|V(T^i)| + |V(T^j)| = |V(T)| - 1$. Now, there are two possible cases - \\
(i) $\C_{T^i} \cup \C_{T^j} = \C_T$ \\
 In this case, $|\B_{T^i} + \B_{T^j}| = |\B_T|$. By Induction hypothesis, we get the upper bound as 
 $$\sum_{i \in V}x_i \leq \frac{|V(T^i)| + |\B_{T^i}| - |\C_{T^i}|}{2} +\frac{|V(T^j)| +  |\B_{T^j}| - |\C_{T^j}|}{2} < \frac{|V(T)| +  |\B_{T}| - |\C_{T}|}{2}.$$
(ii) $\C_{T^i} \cup \C_{T^j} = \C_T \backslash c_j$\\
In this case, $|\B_{T^i} + \B_{T^j}| = |\B_T| - 1$. This is because, when $c_j$ is no longer a center on removing a neighbouring node in $P_{ij}$, it must have had exactly two other branches other than $P_{ij}$. Now, since $P_{ij}$ is not a branch in the new graph, the two other branches are counted only as one along with the node $c_j$ in the branch. By Induction hypothesis, we get the upper bound as 
 $$\sum_{i \in V}x_i \leq \frac{|V(T^i)| + |\B_{T^i}| - |\C_{T^i}|}{2} +\frac{|V(T^j)| +  |\B_{T^j}| - |\C_{T^j}|}{2} < \frac{|V(T)| +  |\B_{T}| - |\C_{T}|}{2}$$.

Hence, the aggregate play is at most $\frac{|V(G)| + |\B_T| - |\C_T|}{2}$ for all equilibria. \hfill $\blacksquare$

\end{subsection}

\subsection{Proof of \cref{thm:treewelapprox}}

 Let the aggregate play maximizing ICE be given by $x^*$. 
 \begin{enumerate}[label = (\alph*)]
  \item We maximize the RHS of \cref{eqn:welbound2} over $x$ to get 
\begin{align}\nonumber
 W(x;\delta,G) &\leq  n(b(\estar) - c\estar) + cd_{max}\delta\sigma_b\sum_{i \in V} x^*_i \\ \label{eqn:maxuppbndtree}
 &\leq n(b(\estar) - c\estar) + cd_{max}\delta\sigma_b\estar\left(\frac{|V(G)| + m - |\C_T|}{2}\right)
\end{align}
Since \cref{eqn:maxuppbndtree} holds for every $x$ such that $\frac{x}{\estar} \in \SOL_\delta(G)$, we have 
\begin{equation*}
 W^*(\delta,G) \leq n(b(\estar) - c\estar) + cd_{max}\delta\sigma_b\estar\left(\frac{|V(G)| +  m - |\C_T|}{2}\right).
\end{equation*}
 \item Since \cref{eqn:welbound1} holds for all $x$, we consider this equation for $x^*$, \ie 
\begin{equation*}
  W^*(\delta,G) \geq W(x^*;\delta,G) \geq n(b(\estar) - c\estar) + c((\delta + 1)\sigma_b - 1)\sum_{i \in V} x^*_i
\end{equation*}
For $\sigma_b \leq \frac{1}{1+\delta}$, to minimize the lower bound, we need to consider the maximum value of $\sum_{i \in V}x^*_i$. Thus, we have for $\sigma_b \leq \frac{1}{1+\delta}$

\begin{equation*}
  W^*(\delta,G) \geq n(b(\estar) - c\estar) + c((\delta + 1)\sigma_b - 1)\estar\left(\frac{|V(G)| + m - |\C_T|}{2}\right).
\end{equation*}
\item For $\sigma_b \geq \frac{1}{1+\delta}$, to minimize the lower bound, we need to consider the minimum value of $\sum_{i \in V}x^*_i$.
Thus, we have for $\sigma_b \geq \frac{1}{1+\delta}$

\begin{equation*}
  W^*(\delta,G) \geq n(b(\estar) - c\estar) + c((\delta + 1)\sigma_b - 1)\estar\left(\frac{|V(G)| + r - |\C_T|}{2}\right),
\end{equation*}
where $r$ is the number of branches which are of odd length.
 \end{enumerate}

\end{section}




\bibliographystyle{plain} 
\bibliography{references}





\end{document}

%% file: macros.tex
\usepackage{amsmath, amssymb, xspace,url}
\usepackage{epsfig}
\usepackage{longtable}
\usepackage{color}

\def\bfone{{\bf 1}}
\def\bfe{{\bf e}}
\def\t{^\top}

\def\bkE{{\rm I\kern-.17em E}}
\def\bk1{{\rm 1\kern-.17em l}}
\def\bkD{{\rm I\kern-.17em D}}
\def\bkR{{\rm I\kern-.17em R}}
\def\bkP{{\rm I\kern-.17em P}}

\def\bkZ{{\bf{Z}}}
\def\bkE{{\rm I\kern-.17em E}}
\def\bk1{{\rm 1\kern-.17em l}}
\def\bkD{{\rm I\kern-.17em D}}
\def\bkR{{\rm I\kern-.17em R}}
\def\bkP{{\rm I\kern-.17em P}}
		\def\bkE{{\rm I\kern-.17em E}}
		\def\bk1{{\rm 1\kern-.17em l}}
		\def\bkD{{\rm I\kern-.17em D}}
		\def\bkR{{\rm I\kern-.17em R}}
		\def\bkP{{\rm I\kern-.17em P}}
		\def\bkY{{\bf \kern-.17em Y}}
		\def\bkZ{{\bf \kern-.17em Z}}
		\def\bkC{{\bf  \kern-.17em C}}

\def\Kscr{{\cal K}}

\def\SOL{{\rm SOL}}

\def\LCP{{\rm LCP}}

\def\NE{{\rm NE}}

\let\forallnew\forall
\renewcommand{\forall}{\forallnew\ }
\let\forall\forallnew

\def\b12{(\beta_1,\beta_2)}

\newcommand{\Real}{\ensuremath{\mathbb{R}}}

\def\half  {{\textstyle{1\over 2}}}

\def\hbar{\skew{4.2}\bar h}

\def\superstar{^{\raise 0.5pt\hbox{$\nthinsp *$}}}
\def\SUPERSTAR{^{\raise 0.5pt\hbox{$*$}}}

\def\lamstarT {\lambda^{\raise 0.5pt\hbox{$\nthinsp *$}T}}

\def\estar{e\superstar}

\newlength{\noteWidth}
\long\def\notes#1{\ifinner
{\tiny #1}
\else
\marginpar{\parbox[t]{\noteWidth}{\raggedright\tiny #1}}
\fi\typeout{#1}}
 \def\notes#1{\typeout{read notes: #1}} 

\newcommand{\ie}{i.e.\@\xspace} 
\newcommand{\etal}{et al.\@\xspace} 

\def\spose#1{\hbox to 0pt{#1\hss}}

\def\text #1{\hbox{\quad#1\quad}}

\makeatletter
\newcommand{\pushright}[1]{\ifmeasuring@#1\else\omit\hfill$\displaystyle#1$\fi\ignorespaces}
\newcommand{\pushleft}[1]{\ifmeasuring@#1\else\omit$\displaystyle#1$\hfill\fi\ignorespaces}
\makeatother

\def\nthinsp{\mskip -2   mu}

%% file: macros-math.tex
\newtheorem{theorem}{Theorem}[section]

\newtheorem{lemma}{Lemma}[section]

\newtheorem{proposition}[theorem]{Proposition}

\newtheorem{corollary}[theorem]{Corollary}

\newtheorem{definition}{Definition}[section]

\newcounter{example}
\renewcommand{\theexample}{\thesection.\arabic{example}}
\newenvironment{examplec}[1][]{\refstepcounter{example}
\par\medskip \noindent%
   \textbf{Example~\theexample. #1} \rmfamily}{\hfill $\square$   \hspace{-4.5pt} \vspace{6pt}}

\newcounter{remark}
\renewcommand{\theremark}{\thesection.\arabic{remark}}

{\begin{list}{}%
         {\setlength{\leftmargin}{#1}}%
         \item[]%
}
{\end{list}}

		\def\bsp{\begin{split}}
		\def\beq{\begin{eqnarray}}
		\def\bal{\begin{align*}}
		\def\bc{\begin{center}}
		\def\be{\begin{enumerate}}
		\def\bi{\begin{itemize}}
		\def\bs{\begin{small}}
		\def\bS{\begin{slide}}
		\def\ec{\end{center}}
		\def\ee{\end{enumerate}}
		\def\ei{\end{itemize}}
		\def\es{\end{small}}
		\def\eS{\end{slide}}
		\def\eeq{\end{eqnarray}}
		\def\eal{\end{align*}}
		\def\esp{\end{split}}
		\def\qed{ \vrule height7.5pt width7.5pt depth0pt}  

	\def\cp2problem#1#2#3#4{\fbox
		 {\begin{tabular*}{0.9\textwidth}
			{@{}l@{\extracolsep{\fill}}l@{\extracolsep{6pt}}l@{\extracolsep{\fill}}c@{}}
				#1 & & $#4 $ 
			\end{tabular*}}}

		\renewcommand{\emph}[1]{\textbf{#1}}

		\def\bkE{{\rm I\kern-.17em E}}
		\def\bk1{{\rm 1\kern-.17em l}}
		\def\bkD{{\rm I\kern-.17em D}}
		\def\bkR{{\rm I\kern-.17em R}}
		\def\bkP{{\rm I\kern-.17em P}}
		
		\def\bkZ{{\bf{Z}}}

\newcommand {\beeq}[1]{\begin{equation}\label{#1}}
\newcommand {\eeeq}{\end{equation}}
\newcommand {\bea}{\begin{eqnarray}}
\newcommand {\eea}{\end{eqnarray}}

\def\texitem#1{\par\smallskip\noindent\hangindent 25pt
               \hbox to 25pt {\hss #1 ~}\ignorespaces}

